\documentclass[preprint,aps,floats,nofootinbib,amssymb,superscriptaddress]{revtex4}
\usepackage[sort&compress]{natbib}
\usepackage{epsf,epsfig}
\usepackage{amsmath}
\usepackage{hyperref}
\usepackage{subfigure}
\usepackage{graphicx}
\usepackage{color}
\usepackage{mathrsfs}

\newcommand{\be}{\begin{equation}}
\newcommand{\ee}{\end{equation}}
\newcommand{\bea}{\begin{eqnarray}}
\newcommand{\eea}{\end{eqnarray}}
\newcommand{\nn}{\nonumber}
\newcommand{\noi}{\noindent}
\newcommand{\ba}{\begin{array}}
\newcommand{\ea}{\end{array}}
\newcommand{\bi}{\begin{itemize}}
\newcommand{\ei}{\end{itemize}}
\newcommand{\proph}{\mathscr{P}_{h}}
\newcommand{\prophp}{\mathscr{P'}_{h}}
\newcommand{\shat}{\hat{s}}
\newcommand{\rts}{\sqrt{\hat{s}}}
\newcommand{\tbar}{\bar{t}}
\newcommand{\bbar}{\bar{b}}
\newcommand{\wplus}{W^{+}}
\newcommand{\wminus}{W^{-}}
\newcommand{\betat}{\beta_{t}}
\newcommand{\ecm}{\sqrt{s}}

\newcommand{\gev}{\; \textrm{GeV}}

\begin{document}

\title{\vspace*{.75in} Azimuthal Correlations in Top Pair Decays and The Effects of New Heavy Scalars}


\author{Vernon Barger}\affiliation{Department of Physics, University of Wisconsin, Madison, Wisconsin 53706, USA}
\author{Wai-Yee Keung}\affiliation{Department of Physics, University of Illinois at Chicago, Chicago, Illinois 60607, USA}
\author{Brian Yencho}\affiliation{Departament d'Estructura i Constituents de la Mat\`eria,
Institut de Ci\`encies del Cosmos (ICCUB), \\
Universitat de Barcelona, Mart\'i Franqu\`es 1, 08028 Barcelona, Spain}

\thispagestyle{empty}

\begin{abstract}
\noi We present the azimuthal correlations in the decay products of the top quark pair created in the process $gg \to t \tbar \to b \bbar \wplus \wminus$ calculated in leading-order QCD.  We then discuss deviations in this distribution that could arise from New Physics contributions, focusing, in particular, on the case of a new heavy scalar coupling to the top quark and produced via the gluon fusion mechanism.  Within this scenario we consider two cases: one in which the top quark is the only particle with a significant coupling to the scalar and another in which there are additional heavy fermions that modify the effective coupling of the scalar to gluons.  We discuss the effects on the overall cross section and azimuthal correlations of the top quark decay products, with particular emphasis on the effect of interference between the QCD and New Physics contributions and on the existence of $CP$-nonconserving couplings.
\end{abstract}

\maketitle

\section{Introduction} \label{sec:intro}

As the heaviest known Standard Model (SM) fermion and one which can now be copiously produced at the CERN LHC, the top quark is an interesting tool in the search for New Physics (NP) beyond the SM.  That the top quark mass $m_{t}$ is so similar to the vacuum expectation value of the SM Higgs boson, $\langle H \rangle = (v / \sqrt{2} ) \approx 174$~GeV, is suggestive of a possible relationship with any New Physics at or near the scale of electroweak symmetry breaking that is now being probed.  Indeed, there has been much theoretical and phenomenological effort recently due to evidence seen at the Fermilab Tevatron of a possible "forward-backward asymmetry" in top pair production \cite{Abazov:2007qb,Aaltonen:2011kc}, with models invoking such possibilities as new $W'$ and $Z'$ bosons \cite{Barger:2010mw,Duraisamy:2011pt,Frank:2011rb}, axi-gluons \cite{Krnjaic:2011ub,Zerwekh:2011wf}, extended Higgs sectors \cite{Wang:2011mr}, new fermions \cite{Davoudiasl:2011tv}, and other exotic scenarios \cite{Cui:2011xy}.

In this paper, we will be concerned not with the angular distributions of the top quarks themselves, but with those of their decay products.  We put particular emphasis on the ability to use these correlations to probe the $CP$ nature of a new scalar resonance, most notably in the case that it is not a $CP$ eigenstate and therefore introduces a new source of $CP$ violation.  There have been many previous studies looking into the effect of resonances of varying spin and $CP$ properties on top pair production, including the impact on the invariant mass spectrum and the behavior of various observables sensitive, in particular, to $CP$ \cite{Gaemers:1984sj,Bernreuther:1993hq,Bernreuther:1993df,Chang:1993jy,Dicus:1994bm,Gunion:1996vv,Bernreuther:1997gs,Bernreuther:1998qv,Accomando:2006ga,Frederix:2007gi,Bernreuther:2008ju,Baumgart:2011wk,He:2011ws}.  Here, we are interested in how the angular distributions of the complete results, including the QCD prediction and its interference with the resonant signal, could be detected as a deviation from the prediction of QCD alone.  We discuss the results as they would appear at the LHC running at a design center-of-mass energy of $\rts=14$~TeV and note the difficulties that arise in both detection of the signal generally and the interpretation of the angular correlations in the $CP$-violating case.

The paper is organized as follows:  In Sec.~\ref{sec:qcd}, we give the results for the azimuthal correlations predicted by leading-order (LO) QCD.  In Sec.~\ref{sec:np}, we describe the models used in our discussion of New Physics contributions.  The effects on the top pair invariant mass distribution are then given and discussed in Sec.~\ref{sec:mtt} and the those of the azimuthal correlations are given in Sec.~\ref{sec:correlations}.  We offer concluding remarks in Sec.~\ref{sec:conclusions}.

\section{QCD Predictions} \label{sec:qcd}

\begin{figure}[tb]
\centering
\subfigure[]{\includegraphics[clip,width=0.40\textwidth]{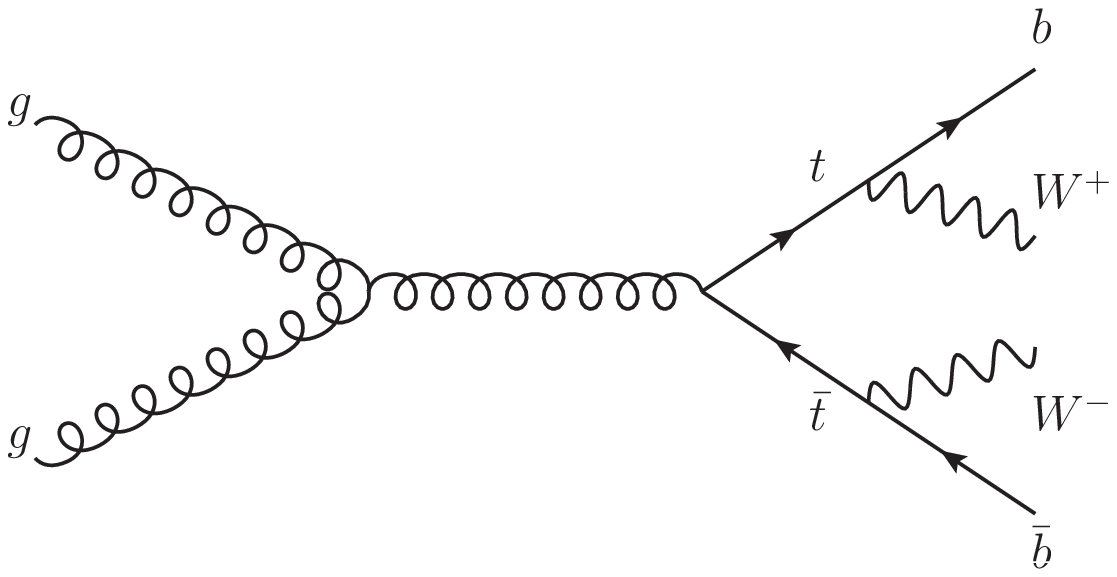}} \\
\subfigure[]{\includegraphics[clip,width=0.30\textwidth]{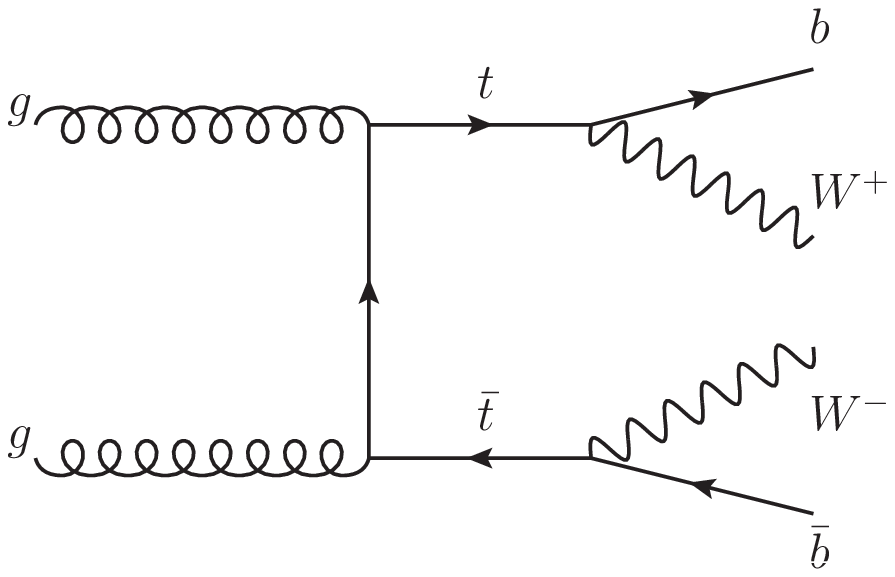}} \hspace{1cm}
\subfigure[]{\includegraphics[clip,width=0.30\textwidth]{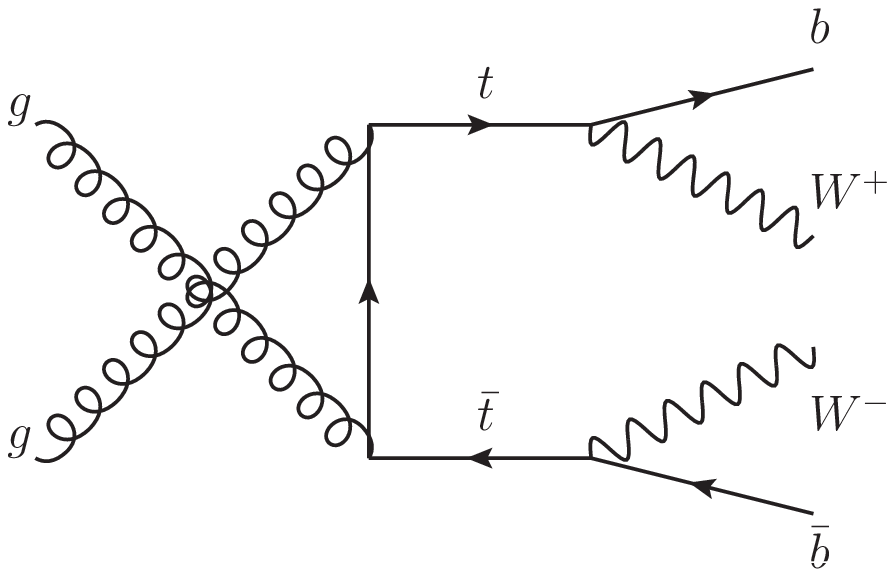}}
\caption{Diagrams contributing to the process $gg \to t\tbar \to b\bbar \wplus \wminus$ at LO in QCD.}
\label{fig:diagrams_qcd}
\end{figure}

The process $gg \to t \tbar \to b \bbar \wplus \wminus$ is given, to leading-order (LO) in QCD, by the three diagrams\footnote{Because of the triple gluon vertex in Fig.~\ref{fig:diagrams_qcd}(a) there may in principle be an additional diagram with the two external gluons replaced by ghosts, depending of the choice of gauge.  We choose a gauge such that this diagram is absent but in which the sum over gluon polarizations is modified.  This choice is used consistently across all calculations throughout the paper.} listed in Figs.~\ref{fig:diagrams_qcd}(a)-\ref{fig:diagrams_qcd}(c).  While $b \bbar \wplus \wminus$ are not the true final-state particles -- the $W$ bosons will decay and the $b$ quarks will hadronize -- we assume that the events can be sufficiently reconstructed to be described at this level.  In practice, this involves significant experimental challenges which vary with the decay modes of the $W$ bosons:  in the fully hadronic channel, one must worry about the large SM background to the overall process $b b j j j j$ in addition to the complicated combinatorics involved with the large number of jets, while in the leptonic modes there is difficulty in precise event reconstruction due to the missing energy carried away by the neutrinos and the unknown components of their momentum along the beam axis (see, e.g., the discussions by the ATLAS and CMS collaborations in Refs.~\cite{Ball:2007zza,Aad:2009wy}). There are ways in which these issues can be partially alleviated using various constraints, such as requiring the reconstruction of the top quark and $W$-boson masses \cite{Frederix:2007gi}, and we note that our results are independent of the particular $W$-boson decay modes and can therefore be applied to any reconstructed experimental sample of $b \bbar \wplus \wminus$.  In what follows, we will assume maximum event reconstruction and selection efficiency for illustrative purposes while noting the above limitations.

\begin{figure}[tb]
\centering
\includegraphics[clip,width=0.60\textwidth]{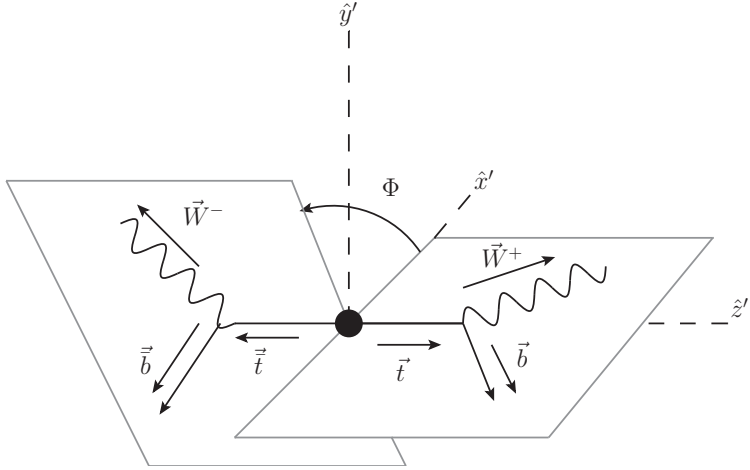}
\caption{This defines the geometry relevant to the measurement of the angle $\Phi$.  In the $t \tbar$ rest frame we identify the $t$ quark 3-momentum as the $\hat{z}'$ direction (which is not to be confused with the beam direction in the laboratory frame, often denoted as $\hat{z}$).  The angle $\Phi$ is then the relative azimuthal angle between the decay planes of the $t$ and $\tbar$ about this axis as indicated.  This angle can also be understood equivalently as the difference in the azimuthal angles of either the $W$ bosons or $b$ quarks about the $\hat{z}'$ axis.}
\label{fig:setup}
\end{figure}

We now specify the frame and overall geometry in which the final results will be interpreted.  We define the $z'$ axis as the $\vec{t}$ direction in the $t \bar{t}$ rest frame.  In the individual top quark rest frames, the $W$ boson 3-momentum vectors (or equivalently the $b$ quark vectors) can be defined relative to this axis by

\bea
\label{eq:momenta}
\frac{\vec{W}_{1}}{|\vec{W}_{1}|} & = & (\sin\theta_{1}\cos\phi_{1},\sin\theta_{1}\sin\phi_{1},\cos\theta_{1}), \\ \nn
\frac{\vec{W}_{2}}{|\vec{W}_{2}|} & = & (\sin\theta_{2}\cos\phi_{2},\sin\theta_{2}\sin\phi_{2},\cos\theta_{2}), \\ \nn
\eea

\noi where here $W_{1} = W^{+}$ and $W_{2} = W^{-}$.  The angles $\phi_{1}$ and $\phi_{2}$ are invariant under boosts along the $\hat{z}'$ direction.  In what follows we will be interested in the relative azimuthal angle of the $W$ bosons / $b$ quarks in the $t \tbar$ frame about the $z'$ axis and therefore define the angle $\Phi = (\phi_{2}-\phi_{1})$.  This angle represents the azimuthal separation of the two decay planes as indicated in Fig.~\ref{fig:setup}.  We note that although we have indicated that the $W^{+}$/$W^{-}$ bosons and the $b$/$\bbar$ quarks are distinguishable from one another, this could in principle not be the case experimentally.  However, the results that follow still hold when the proper identification of states is made.  What is relevant is that for each event a $W$ boson and $b$ quark can be paired uniquely to reconstruct a single top quark.  In this case, we can designate either reconstructed top quark as $t_{1}$, define $\hat{z}' \equiv \hat{t}_{1}$, and associate $W_{1}$ with the $W$ boson resulting from the decay of $t_{1}$.  The relevant angles are then once again defined according to Eq.~(\ref{eq:momenta}) and displayed in Fig.~\ref{fig:setup} with the necessary replacements.

We calculate the full matrix element directly for the process $gg \to t\tbar \to b \bbar \wplus \wminus$ using the diagrams shown in Figs.~\ref{fig:diagrams_qcd}(a)-\ref{fig:diagrams_qcd}(c) assuming real, on-shell $W$ bosons.  In addition, we make use of the narrow width approximation for the top quark propagators and integrate over all angles except $\Phi = (\phi_{2}-\phi_{1})$ to find the normalized, parton-level result for the distribution in $\Phi$:

\be
\label{eq:dist_qcd}
\frac{1}{\hat{\sigma}_{\rm QCD}} \frac{d \hat{\sigma}_{\rm QCD}}{d \Phi} = 
\left(\frac{1}{2\pi}\right)
\left[ 1 + 
C_{\rm QCD}(\beta_{t}) 
\left(\frac{\pi}{4}\right)^{2} 
\frac{\left(1 - 2 \rho_{w} \right)^{2}}{\left(1 + 2 \rho_{w} \right)^{2}}
\cos(\Phi)
\right],
\ee

\noi where $\beta_{t} = \sqrt{1-4 m_{t}^{2}/ \shat}$, $\rho_{w} = (M_{W}/m_{t})^{2}$, and the coefficient is defined as

\be
\label{eq:c_qcd}
C_{\rm QCD}(\beta_{t}) = \left(\frac{(1-\beta_{t}^{2})}{\beta_{t}^{2}}\right)
\left(
\frac{
\beta_{t}(33-31\beta_{t}^{2})-(1-\beta_{t}^{2})(33-2\beta_{t}^{2})\tanh^{-1}(\beta_{t})
}
{\beta_{t}(59-31\beta_{t}^{2})-2(33-18\beta_{t}^{2}+\beta_{t}^{4})\tanh^{-1}(\beta_{t})
}
\right) .
\ee

\begin{figure}
\subfigure[]{\includegraphics[clip,width=0.45\textwidth]{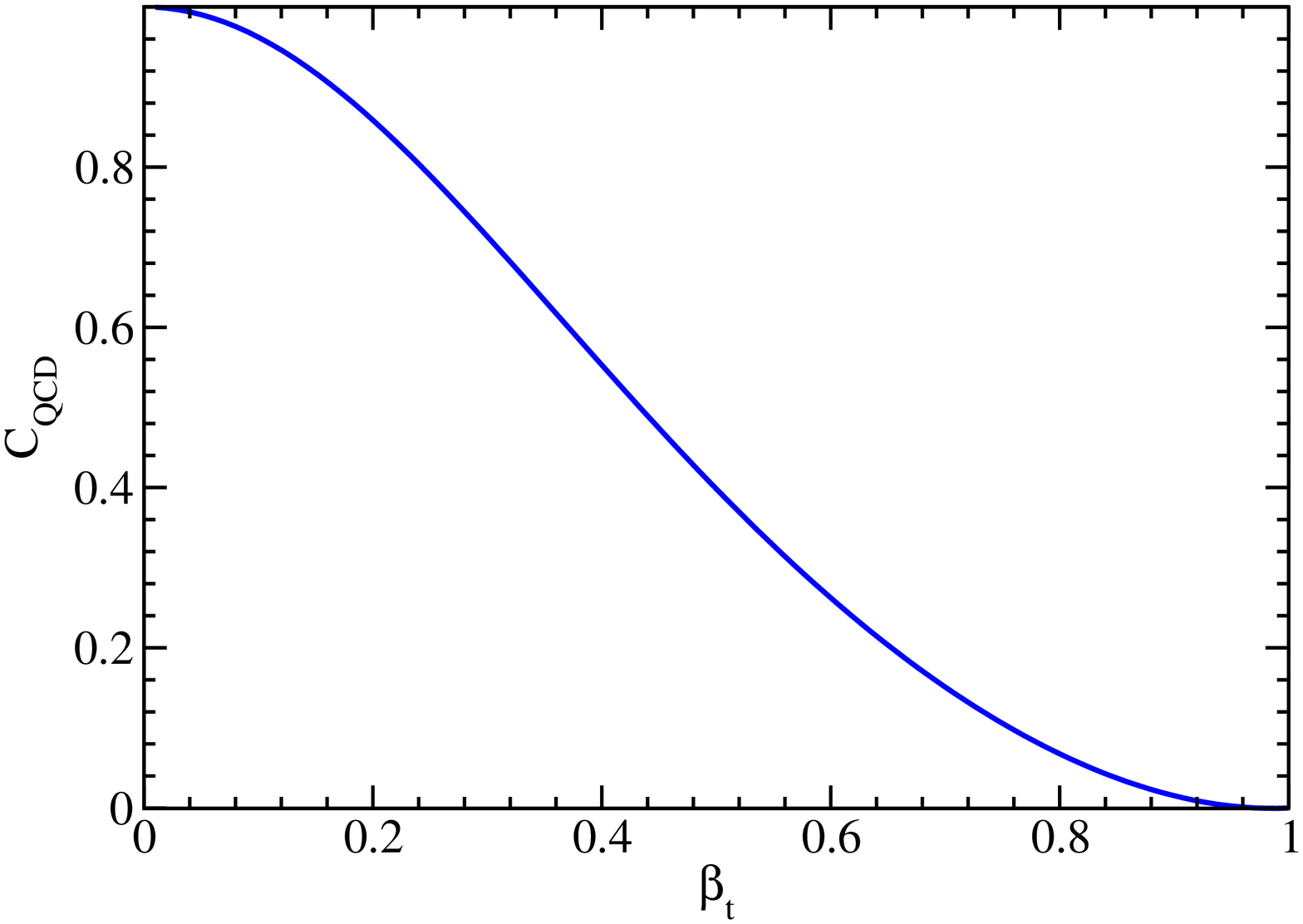}}
\subfigure[]{\includegraphics[clip,width=0.45\textwidth]{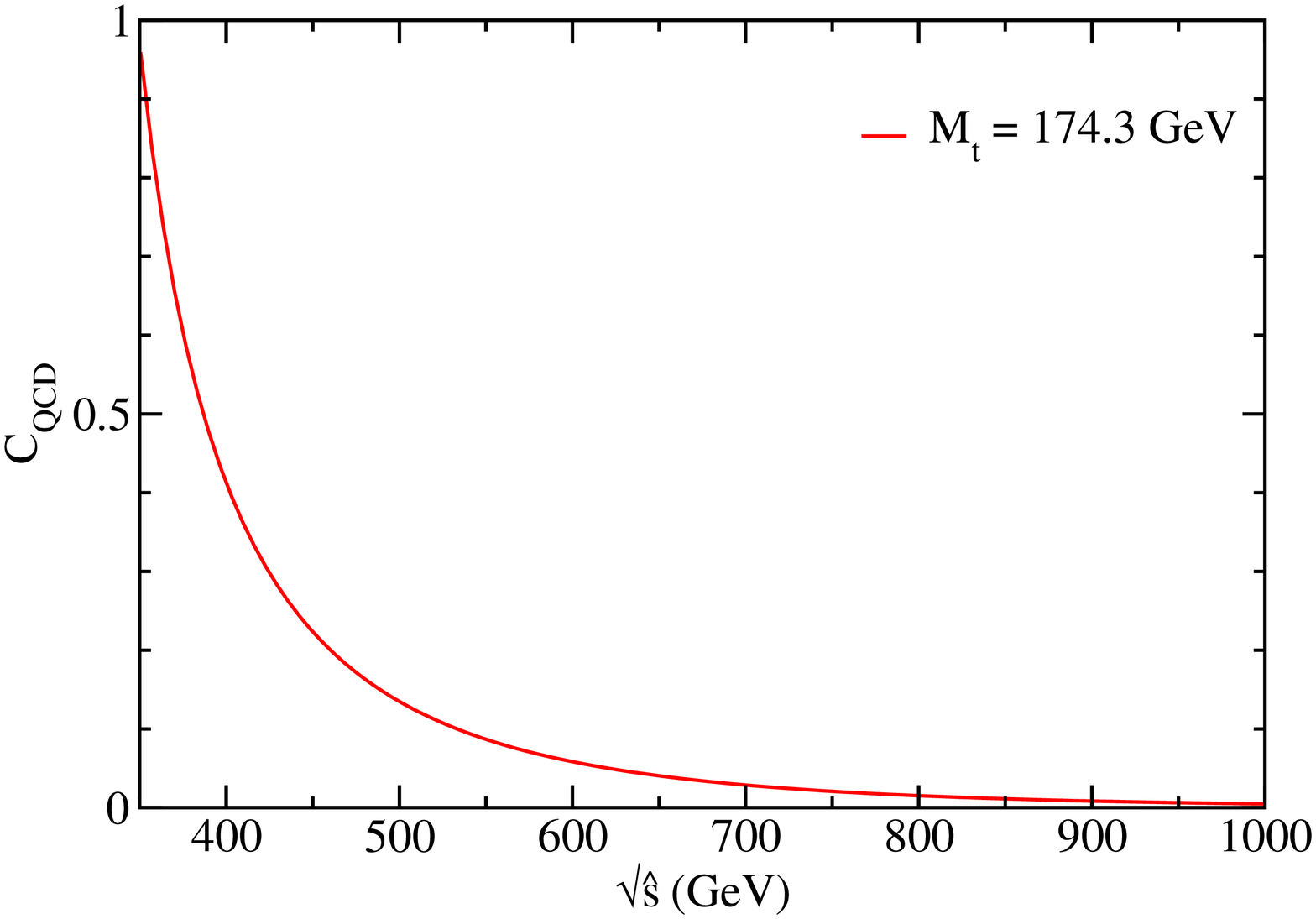}}
\caption{Amplitude factor $C_{\rm QCD}$ as a function of (a) $\beta_{t}$ and (b) $\rts$.}
\label{fig:amplitude}
\end{figure}

\noi The overall partonic level cross section is given by

\bea
\label{eq:sigma_qcd}
\hat{\sigma}_{\rm QCD} \left(\shat\right) & =  &\Bigg[ \left(\frac{\alpha_{s}^{2} \pi}{48}\right) \left(\frac{1}{\hat{s}}\right) \big[
2\left( 33 -18 \beta_{t}^{2} + \beta_{t}^{4} \right) \tanh^{-1}(\beta_{t})
-\beta_{t} \left(59-31\beta_{t}^{2}\right)
\big]  \\ \nn
& \hspace{0.5cm} & \times \textrm{BR}(t \to b W)^{2}
\Bigg],
\eea

\noi where BR$(t \to b W) \equiv $ BR$(t \to b \wplus)$ = BR$(\tbar \to \bbar \wminus)$ is the top quark branching fraction.

We plot the coefficient $C_{\rm QCD}(\beta_{t})$ in Figs.~\ref{fig:amplitude}(a) and \ref{fig:amplitude}(b) as a function of both $\betat$ and $\rts$.  In the latter, we take a top mass of $m_{t} = 174.3\gev$, which we use throughout the paper.  As is evident from this figure, it is a well-behaved function which takes the limits

\be
C_{\rm QCD} \to \left\{
\begin{array}{ccc}
1  & \textrm{as} & \betat \to 0 \\
0  & \textrm{as} & \betat \to 1 \\
\end{array}
\right. .
\ee

\noi The $\cos\Phi$ dependence of the distribution is therefore maximal right above the production threshold of the $t\tbar$ pair and then rapidly goes away at higher energies.  This behavior stems from the different energy dependences of the relative contributions to the total amplitude coming from initial state gluons of varying helicity, as discussed in Ref.~\cite{Mahlon:2010gw}.

\begin{figure}
\includegraphics[clip,width=0.75\textwidth]{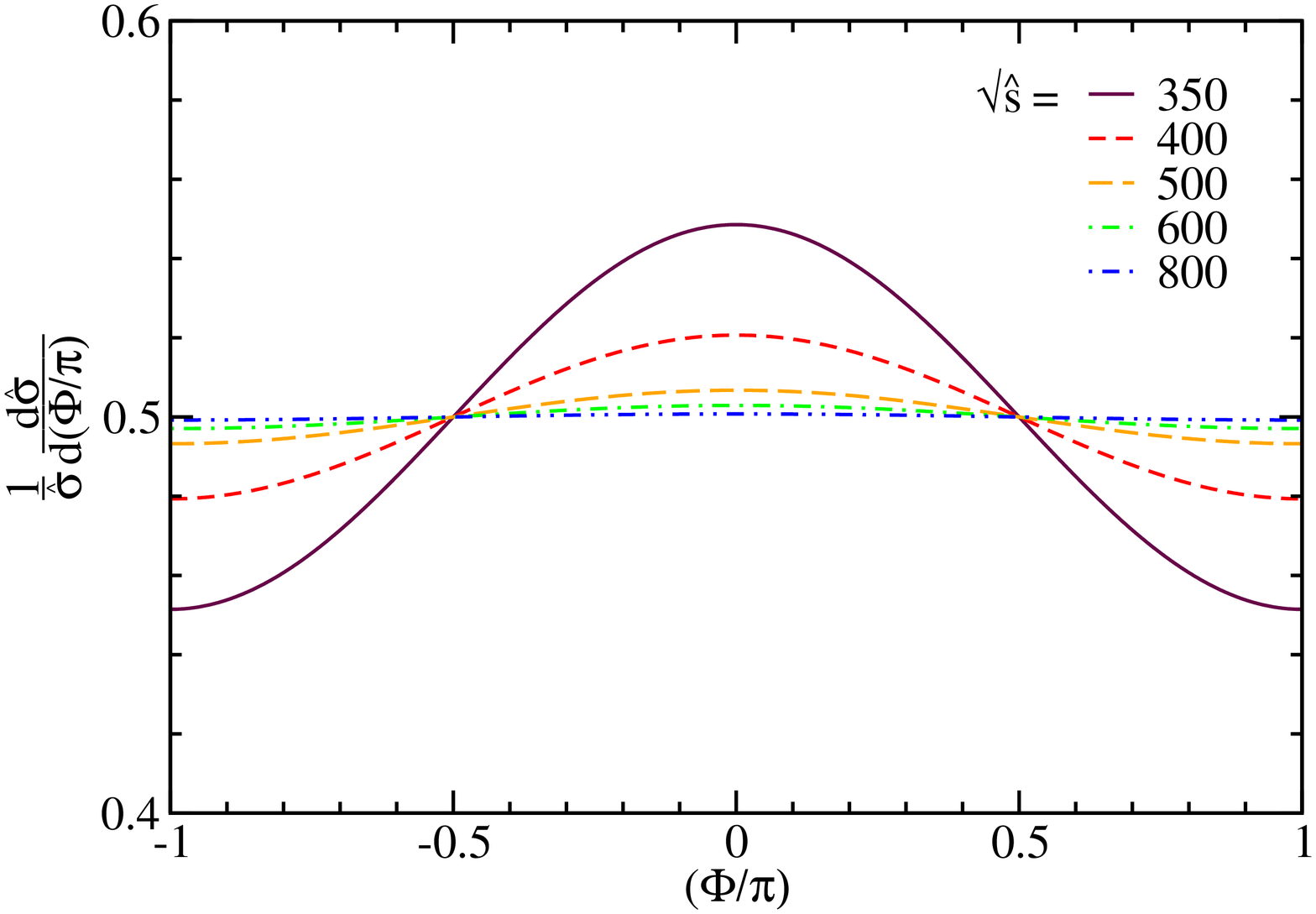}
\caption{Normalized partonic level $\Phi$ distribution for several different values of the center-of-mass energy $\rts$ predicted by LO QCD for the process $gg \to t \tbar \to b \bbar \wplus \wminus$.}
\label{fig:phi_qcd}
\end{figure}

We plot the resulting overall behavior of Eq.~(\ref{eq:dist_qcd}) in Fig.~\ref{fig:phi_qcd} for several different values of partonic center-of-mass energy $\rts$.  The strong dependence on $\rts$ is clearly demonstrated, with the distribution appearing nearly flat by $\rts \sim 800$~GeV.  What we are now interested in is how contributions from New Physics may introduce deviations into this behavior.

\section{New Physics Model} \label{sec:np}

As a simple and illustrative example, we consider a scalar particle\footnote{While we suggestively label the state as $h$, it is not necessarily associated with the Higgs boson of the Standard Model, as we intend our results to be more general and model independent.  This choice is simply motivated by the fact that much of the formalism has been borrowed from previous studies more directly applicable to a light, SM-like Higgs boson.} $h$ coupling to the top quark according to the following Lagrangian:

\bea
\label{eq:htt}
\mathscr{L}  & \supset &  - \left(\frac{m_{t}}{v}\right) h \, \bar{t} (A P_{L} + A^{*} P_{R}) t \\ \nn
                      & \supset &  - \left(\frac{m_{t}}{v}\right) h \, \bar{t} ( A_{R} - i A_{I} \gamma^{5} ) t \, ,
\eea

\noi where

\be
\label{eq:a}
A = (A_{ R} + i A_{I}) = |A| e^{i \alpha} \, .
\ee

\noi We restrict $\alpha$ to the range $-\frac{\pi}{2} \le \alpha \le \frac{\pi}{2}$ without loss of generality.  This gives a completely scalar (pseudoscalar) coupling when $\alpha = 0$ ($\pm \frac{\pi}{2}$).  In particular, when $|A|=1$ and $\alpha=0$, this is the same as the top quark coupling to the SM Higgs boson.  Other values of $\alpha$, however, would indicate $CP$ violation.  Such couplings can arise, for example, from multi-Higgs models with complex couplings after rewriting the Lagrangian in the mass basis \cite{Lee:1973iz,Weinberg:1976hu,Branco:1985aq,Khater:2003wq} and have been studied in a variety of contexts \cite{Bernreuther:1993hq,Bernreuther:1993df,Chang:1993jy,Dicus:1994bm,Gunion:1996vv,Bernreuther:1997gs,Bernreuther:1998qv,Lee:2003nta,Accomando:2006ga,Frederix:2007gi,Bernreuther:2008ju,Baumgart:2011wk,He:2011ws}.  Here we are primarily interested in probing the nature of such couplings using the angular correlations described above.

\begin{figure}[tb]
\centering
\subfigure[]{\includegraphics[clip,width=0.40\textwidth]{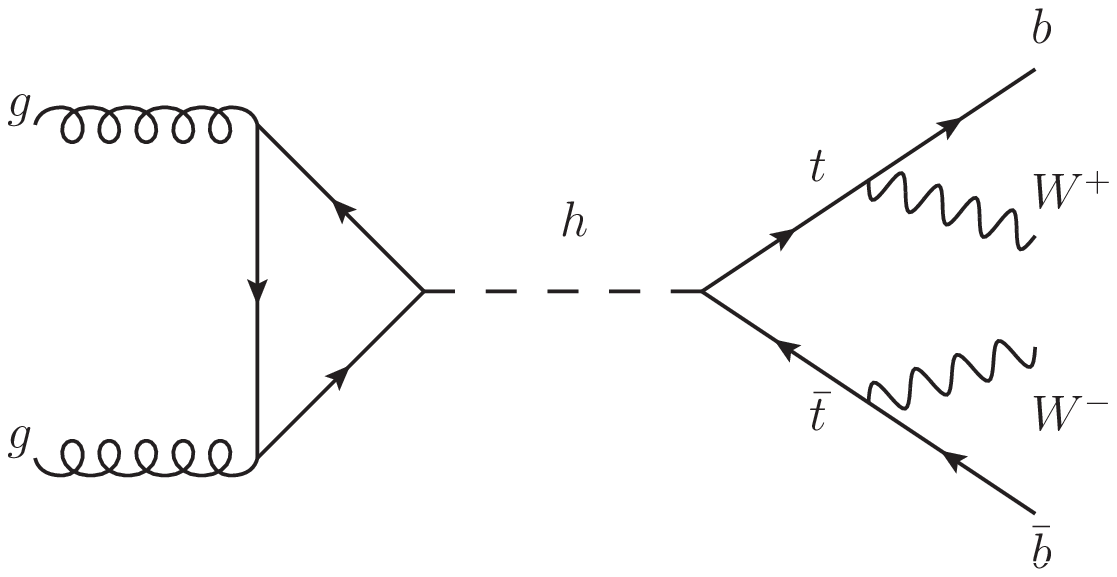}} \hspace{1cm}
\subfigure[]{\includegraphics[clip,width=0.40\textwidth]{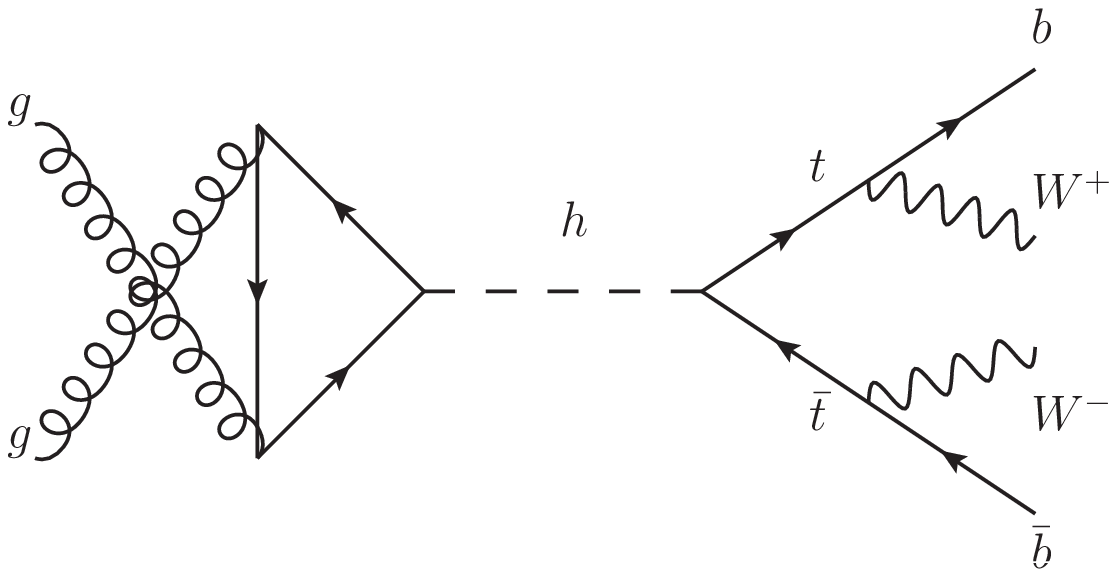}} \\
\caption{Diagrams contributing to the process $gg \to h\to t\tbar \to b\bbar \wplus \wminus$.  Any colored fermions which couple to the new scalar boson $h$ may appear in the fermion loop.}
\label{fig:diagrams_h}
\end{figure}

We assume this scalar is heavy enough to decay on shell to a pair of top quarks.  The existence of this scalar, then, introduces relevant new diagrams to the process $gg \to t\tbar \to b \bbar \wplus \wminus$.  These are given in Figs.~\ref{fig:diagrams_h}(a) and \ref{fig:diagrams_h}(b).  These diagrams can interfere quite strongly with those of the purely QCD process, a point noted by several authors when considering the effects of heavy scalars on $gg \to t\tbar$ production \cite{Dicus:1994bm,Bernreuther:1998qv}.  We will take special care to account for this interference analytically when discussing the complete azimuthal correlations below.

Because of the scalar nature of $h$, the production process $gg \to h^{(*)}$ can be calculated separately and summed between the two contributing diagrams to give the amplitude:

\bea
\label{eq:amp_ggh}
i \mathcal{M}(gg \to h^{(*)} ) = \left( i \delta^{a b} \right)& \Bigg[ &
\left( \frac{\alpha_{s}}{3 \pi v} \right) 
\left( A_{R} \sum_{Q} A_{H}(\tau_{Q}) \right)
\left(-g^{\mu\nu} (p_{1} \cdot p_{2}) + p_{2}^{\mu}p_{1}^{\nu} \right) \\ \nn
&  +  & \left( \frac{\alpha_{s}}{2 \pi v} \right)
\left( A_{I} \sum_{Q} A_{A}(\tau_{Q}) \right)
\left( \epsilon^{\mu\nu\rho\sigma} p_{1\rho}p_{2\sigma}\right)
\Bigg]  \epsilon_{\mu}(p_{1}) \epsilon_{\nu}(p_{2}) ,
\eea

\noi where $\tau_{Q} = (4 m_{Q}^{2} / \hat{s}) $ and

\bea
A_{H}(\tau) & = & \left( \frac{3}{2}\right) \tau \left(1 + (1-\tau) f(\tau) \right) ,\\ \nonumber
A_{A}(\tau) & = & \tau f(\tau) ,\\ \nonumber
f(\tau) &= & \left\{
\begin{array}{ccl}
\arcsin^{2}\left( \frac{1}{\sqrt{\tau}}\right) & \textrm{if} & \tau \ge 1 \\
-\frac{1}{4}\left[ 
\ln\left( \frac{1+\sqrt{1-\tau}}{1-\sqrt{1-\tau}} \right) - i \pi
\right]^{2}
& \textrm{if} & \tau < 1
\end{array}
\right. .
\eea

\noi The functions $A_{H}(\tau)$ and $A_{A}(\tau)$ are normalized such that $A_{H}(\tau), A_{A}(\tau) \to 1$ in the limit $\tau \to \infty$ and are well known from searches of $CP$-even and $CP$-odd Higgs bosons produced via gluon fusion (see, for example, Ref.~\cite{Gunion:1989we} and references therein).  For convenience we will now define

\bea
\label{eq:fhfa}
F_{H} & \equiv  & \sum_{Q} A_{H}(\tau_{Q}) ,\\ \nn
F_{A} & \equiv  & \sum_{Q} A_{A}(\tau_{Q}) .
\eea

\noi The sums appearing here and in Eq.~(\ref{eq:amp_ggh}) run over all the quarks that may contribute to the loop.  It is assumed that any other quarks coupling to the scalar would do so according to Eq.~(\ref{eq:htt}) with the replacement $m_{t} \to m_{Q}$.  This also makes the assumption that the factor $A$ is universal for all heavy fermions.  While not necessarily generic, this will simplify the discussion below.  We also note that in the above calculation, we have assumed that $h$ could in principle be off shell such that it is not necessarily true that $\shat =  m_{h}^{2}$.  Therefore, we use $\tau_{Q} = (4 m_{Q}^{2} / \shat)$ rather than the more commonly used value of $\tau_{Q} = (4 m_{Q}^{2} / m_{h}^{2})$ seen in the literature.

For completeness, we note that in the large $\tau$ limit, the effective coupling of the scalar to two gluons can be characterized by the effective Lagrangian term

\be
\label{eq:eff_lag}
\mathscr{L} \supset -\frac{1}{4} g_{Hgg} \, h \, G^{\mu \nu}_{a} G_{\mu \nu}^{a} -\frac{1}{4} g_{Agg} \,  h \, G^{\mu \nu}_{a} \tilde{G}_{\mu \nu}^{a} \, ,
\ee

\noi where $\tilde{G}_{\mu \nu}^{a} = \frac{1}{2} \epsilon_{\mu\nu\rho\sigma} G^{\rho \sigma, a}$ is the dual to the gluon field strength tensor and 

\bea
\label{eq:eff_coup}
g_{Hgg} & = & -\left( \frac{\alpha_{s}}{3 \pi v} \right) A_{R} F_{H} \, ,\\ \nonumber
g_{Agg} & = & -\left( \frac{\alpha_{s}}{2 \pi v} \right) A_{I} F_{A} \, .
\eea

\noi We note here that we use the definition $\epsilon^{0123}= (- \epsilon_{0123} ) = +1$, which implies that the relation

\be
\textrm{tr}[\gamma^{\mu} \gamma^{\nu} \gamma^{\rho} \gamma^{\sigma} \gamma^{5}] = - 4 i \epsilon^{\mu\nu\rho\sigma}
\ee

\noi was used in the calculation of $g_{Agg}$.  The overall and relative signs of the two terms in Eq.~(\ref{eq:eff_lag}) [when substituting in the values from Eq.~(\ref{eq:eff_coup})] are often unimportant to the studies of Higgs searches.  However, we will be concerned with the interference with the purely QCD process and therefore the proper signs are critical.  Furthermore, when dealing with the case of $CP$ violation -- in which neither $A_{R}$ nor $A_{I}$ are zero -- both terms will make a contribution to this interference.   For this reason we emphasize our use of $\epsilon^{0123}=+1$, as a different choice would change the sign of $g_{Agg}$ in Eq.~(\ref{eq:eff_coup}) and of the $CP$-odd term in Eq.~(\ref{eq:amp_ggh}).  

We will consider two scenarios within this overall framework.  In the first, we consider the top quark to be the only fermion coupling to the new scalar.  The top quark is then the only colored fermion appearing in the loops in Fig.~\ref{fig:diagrams_h} and in the sums in Eqs.~(\ref{eq:amp_ggh}) and (\ref{eq:fhfa}).  We refer to this as the "top-only" scenario. 

The other scenario we consider is one motivated by the possible existence of a fourth generation of fermions.  There are then fourth-generation quarks that could also run in the fermion loops along with the top quark.  In addition to the boost in the production cross section these fermions would provide \cite{Gunion:1989we}, such a scenario could also relieve the tension a heavy scalar may have with electroweak precision observables if it were to be associated with the Higgs boson \cite{Kribs:2007nz,Holdom:2009rf}.  For our purposes, we will only assume for this scenario the existence of two additional quarks with $m_{Q} = 500$~GeV but will, for simplicity, refer to this as the "fourth-generation" scenario.  This particular value of $m_{Q}$ is chosen not only for consistency with direct search bounds on fourth-generation quarks -- with currently published results placing a limit of $m_{b'}\gtrsim 350$~GeV \cite{Aaltonen:2009nr,Chatrchyan:2011em} and preliminary results indicating an exclusion $m_{t'}\gtrsim450$~GeV \cite{Luk:2011np}-- but also to illustrate the effects of contributions to Eq.~(\ref{eq:fhfa}) with $\tau_{Q} > 1$ and therefore which provide contrasting behavior to the top-only scenario, where $\tau_{t} \lesssim 1$.  We note that values of $m_{Q}$ much larger than this could run into problems with perturbative unitarity of the scalar sector \cite{Chanowitz:1978uj,Chanowitz:1978mv}.

In both scenarios, we assume that $h \to t \tbar$ is the only decay mode with an appreciable width, i.e., the partial widths $h \to gg$ and $h \to \gamma \gamma$ are ignored and any new colored fermions are assumed to have $m_{Q} > m_{h}/2$.  The total scalar decay width is then given by

\be
\label{eq:hwidth}
\Gamma_{h} =  \left( \frac{3 m_{h}}{8 \pi} \right) \left(\frac{m_{t}}{v} \right)^{2} \left( A_{R}^{2} \beta_{t}^{2} + A_{I}^2 \right) \beta_{t} \, .
\ee

\noi Unless otherwise stated, we fix the magnitude of $A$ from Eq.~(\ref{eq:a}) to be $|A|=1$.  We note that the observability of the scalar signal depends very strongly on the above assumption.  If, for instance, the scalar is treated as a very heavy Standard Model Higgs boson with no enhancement of the cross section, the implied increase in the scalar decay width from direct couplings to weak gauge bosons would wash out any observable effects \cite{Bernreuther:2008ju}.  The width above represents the smallest possible value for a scalar coupling to the top according to Eq.~(\ref{eq:htt}) and therefore can be thought of as a limiting scenario.

\section{Top Pair Invariant Mass Distributions} \label{sec:mtt}

Prior to our discussion of the effects such models have on the azimuthal correlations of the decay planes of the top quarks, we will first discuss the relative contributions to the invariant mass of the top pair coming from QCD, the scalar signal, and the interference between them.  The primary reason for this is that the deviations from the QCD description of the azimuthal correlations can only in principle be observed in small regions of the top pair invariant mass where the signal is localized.  Furthermore, the size of the deviations will depend on the relative strength of the signal to the QCD background and therefore it is important to first have a thorough understanding of this distribution before proceeding.

At LO, the parton-level cross section for $gg \to h \to t \tbar \to b \bbar \wplus \wminus$ and its interference with QCD are given by

\bea
\label{eq:sigma_h}
\hat{\sigma}_{\rm H}(\shat)  = 
& \Bigg[ &
\left(\frac{\alpha_{s}^{2}}{1536 \pi^{3}}\right)
\left(\frac{m_{t}}{v}\right)^{4} 
(\shat)
|\proph|^{2}
\left( 4 |A_{R} F_{H}|^{2} + 9 |A_{I} F_{A}|^{2} \right) \\ \nn
& \times & \frac{\beta_{t}}{(1-\beta_{t}^{2})}
\left(
A_{R}^{2} \beta_{t}^{2} + A_{I}^{2}
\right)
\times \textrm{BR}(t \to b W)^{2}
\Bigg] \, ,
\eea

\bea
\label{eq:sigma_int}
\hat{\sigma}_{\rm INT}(\shat) = 
- & \Bigg[ & 
\left(\frac{\alpha_{s}^{2}}{96 \pi}\right) \left(\frac{m_{t}}{v}\right)^{2} \tanh^{-1}(\beta_{t})
\left( 2 A_{R}^{2} \beta_{t}^{2} C_{H} + 3 A_{I}^{2} C_{A} \right) \\ \nn
& \times & \textrm{BR}(t \to b W)^{2} 
\Bigg] \, ,
\eea

\noi where we define

\bea
\label{eq:chca}
C_{H} & = & \textrm{Re}\left(F_{H} \prophp \right) \\ \nn
C_{A} & = & \textrm{Re}\left(F_{A} \prophp \right)
\eea

\noi with $F_{H}$ and $F_{A}$ from Eq.~(\ref{eq:fhfa}) and denote the Higgs propagator by

\be
\proph = i \prophp = \frac{i}{(\hat{s}-m_{h}^{2}) + i m_{h} \Gamma_{h}} \, .
\ee

\noi These are calculated in the narrow width approximation for the top quark propagators and are in agreement with previous calculations in the $CP$-conserving limits \cite{Dicus:1994bm}.  The term $\prophp$, which appears in Eq.~(\ref{eq:chca}), is not the full scalar propagator, but the scalar propagator stripped of an overall factor of $i$.  It is important to note that unlike in the signal cross section in Eq.~(\ref{eq:sigma_h}), this factor appears in the interference term in Eq.~(\ref{eq:sigma_int}) unsquared via the functions $C_{H}$ and $C_{A}$.  This results in $\sigma_{\rm INT}(\shat)$ switching signs near, but not typically at, $\shat = m_{h}^{2}$.  This is an important effect in the overall results, as the magnitude of the interference term can be rather large relative to that of the signal.  The resulting negative contributions can give rise to a "peak-dip" structure in the top pair invariant mass distribution, whereby the final result may lie both above and below the QCD prediction with comparable magnitude \cite{Gaemers:1984sj,Dicus:1994bm,Bernreuther:1998qv}.  

The total parton-level cross section for $gg \to t \bar{t} \to b \bar{b} W^{+} W^{-}$ is the sum 

\be
\label{eq:sigma_tot}
\hat{\sigma}_{tot} = \hat{\sigma}_{\rm QCD} + \hat{\sigma}_{\rm H} + \hat{\sigma}_{\rm INT} \, .
\ee

\noi Each of the formulas in Eqs.~(\ref{eq:sigma_qcd}), (\ref{eq:sigma_h}), (\ref{eq:sigma_int}), and  (\ref{eq:sigma_tot}) gives the results for the process $gg \to t \tbar$ (absent the decay of the top quarks) with the removal of the BR$(t \to b W)$ factors.  We can then integrate over a set of parton distribution functions (PDFs) to find the invariant mass distribution of the top pair at the LHC using

\be
\frac{d\sigma}{d\tau} =  \int dy f_{g/p}\left(\sqrt{\tau} e^{y}, \mu_{F}  \right) f_{g/p}\left(\sqrt{\tau} e^{-y}, \mu_{F}  \right) \hat{\sigma} (\tau) \, ,
\ee

\noi where $f_{g/p}$ is the gluon PDF, $\mu_{F}$ is the factorization scale, $y$ is the rapidity of the center-of-mass system in the lab frame, and $\tau = (\hat{s}/s)$.  Then

\be
\frac{d\sigma}{dM_{t\tbar}}  = \frac{d\sigma}{d\sqrt{\hat{s}}} = \left(\frac{2 \sqrt{\hat{s}}}{s}\right) \times  \frac{d\sigma}{d\tau} \left(\tau = (\shat / s) \right) \, .
\ee

\noi The total cross sections are then given by

\be
\sigma = \int dy d\tau f_{g/p}\left(\sqrt{\tau} e^{y}, \mu_{F}  \right) f_{g/p}\left(\sqrt{\tau} e^{-y}, \mu_{F}  \right) \hat{\sigma} (\tau) \, .
\ee

\begin{figure}
\begin{minipage}{0.40\textwidth}
\subfigure[]{\includegraphics[clip,width=1.00\textwidth]{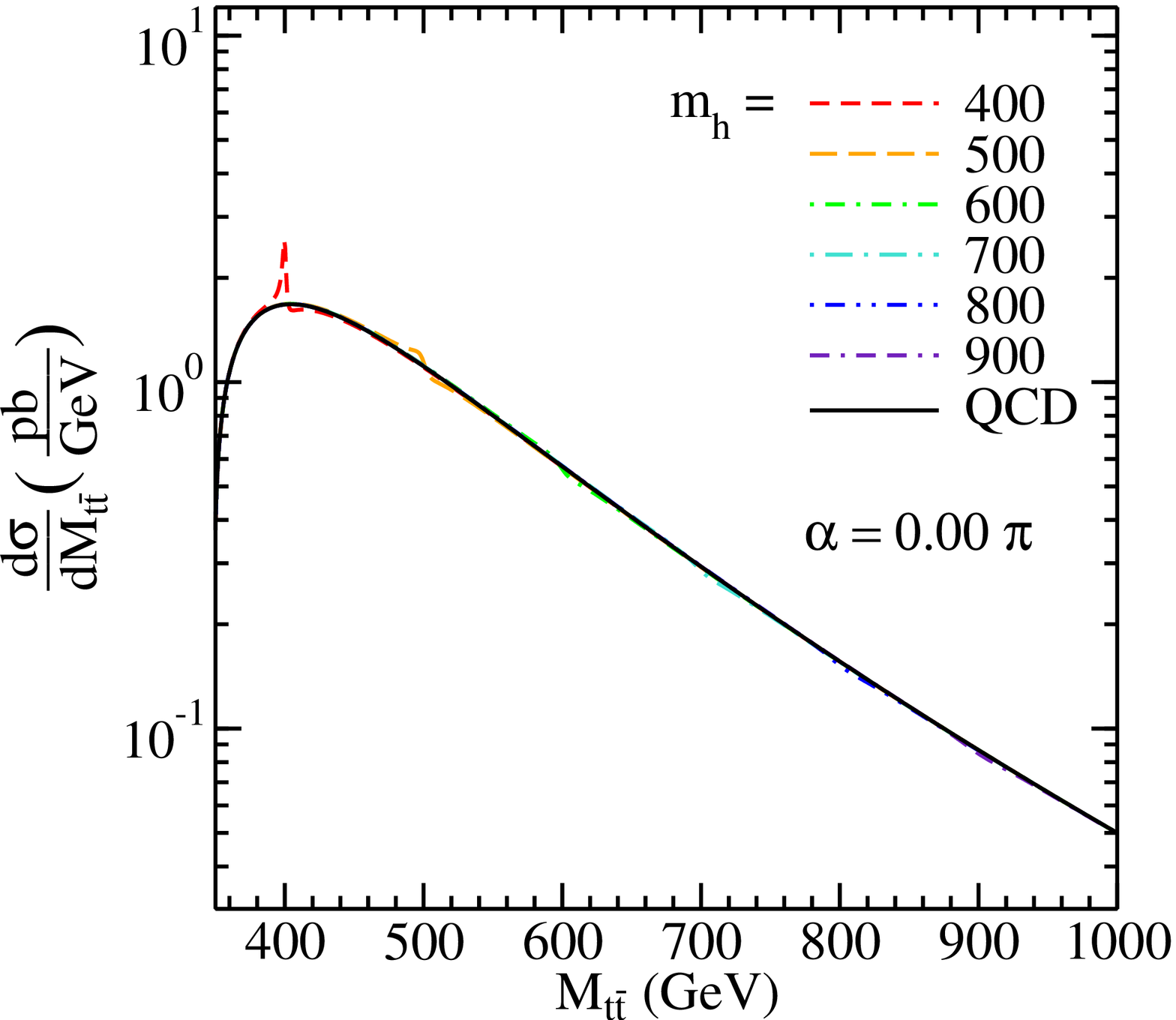}} \\
\subfigure[]{\includegraphics[clip,width=1.00\textwidth]{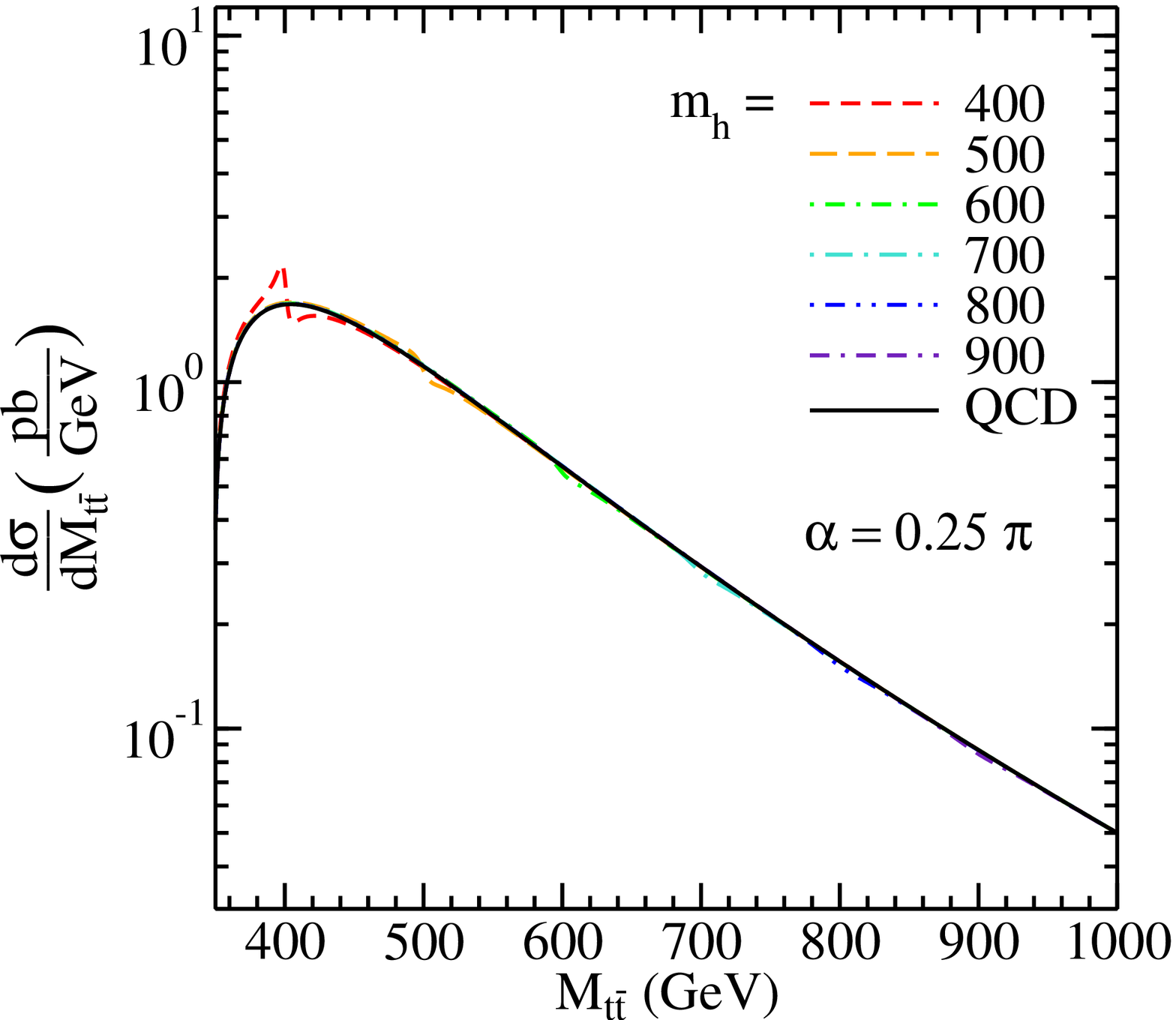}} \\
\subfigure[]{\includegraphics[clip,width=1.00\textwidth]{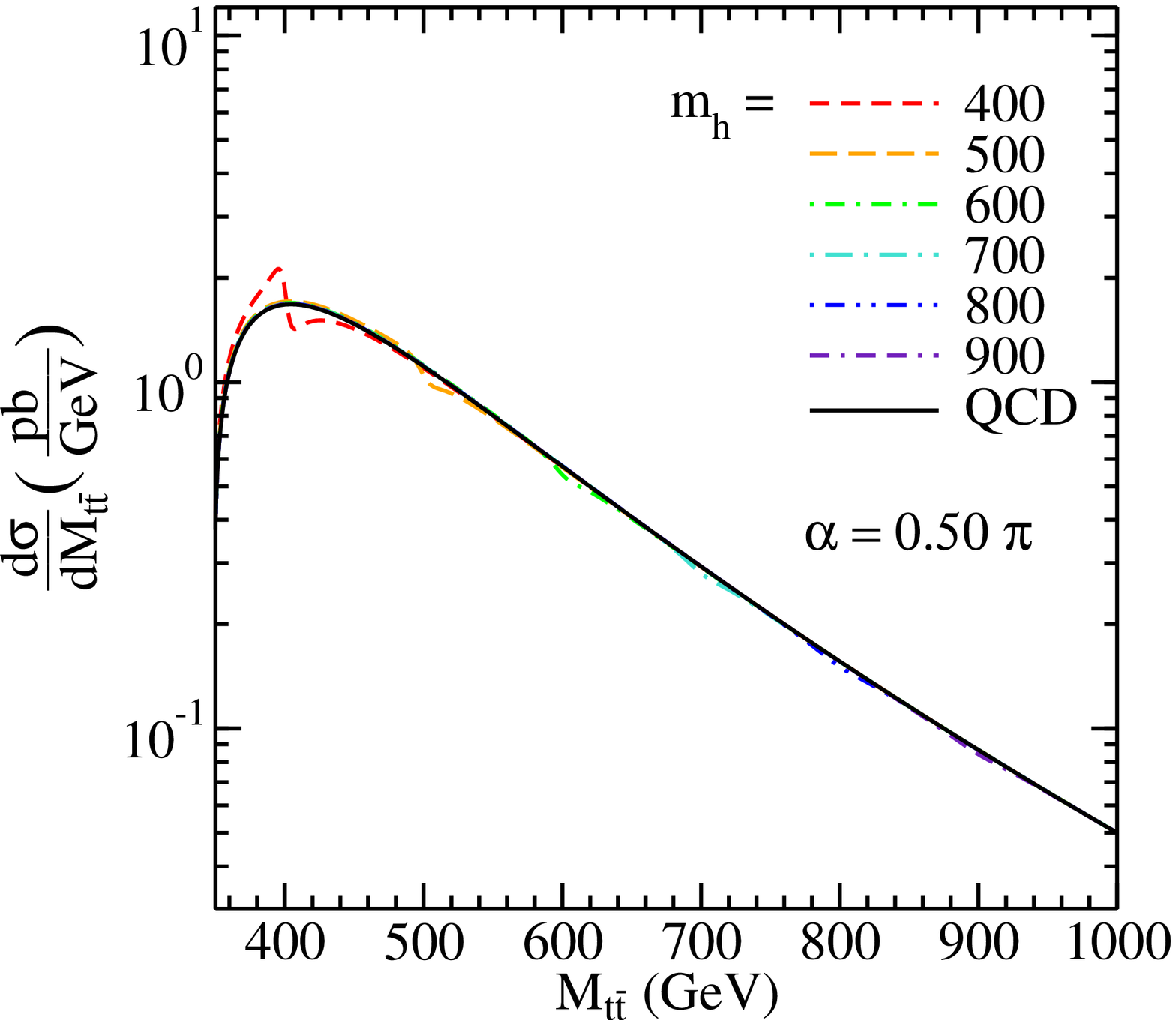}} 
\end{minipage} \hspace{1cm}
\begin{minipage}{0.40\textwidth}
\subfigure[]{\includegraphics[clip,width=1.00\textwidth]{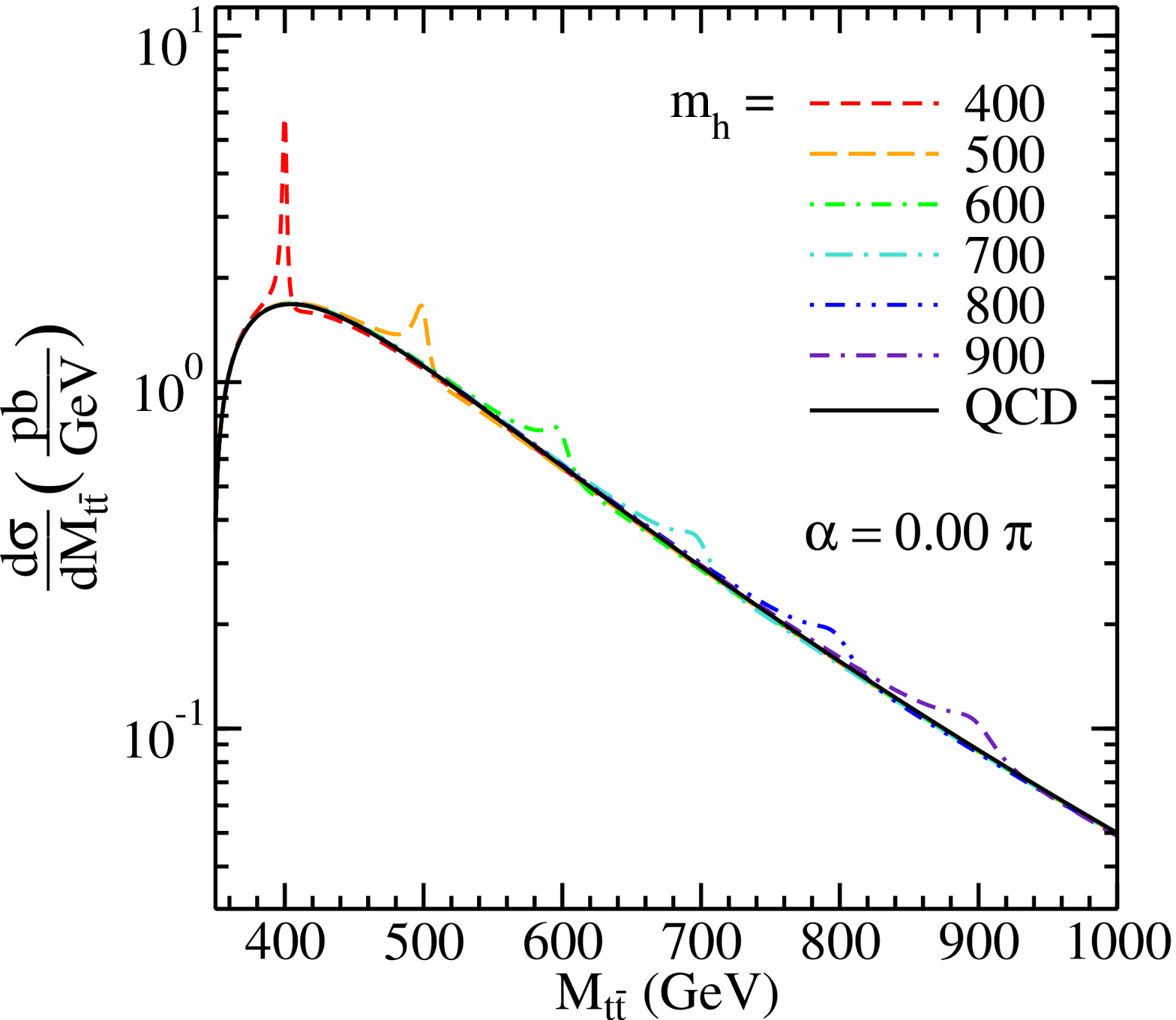}} \\
\subfigure[]{\includegraphics[clip,width=1.00\textwidth]{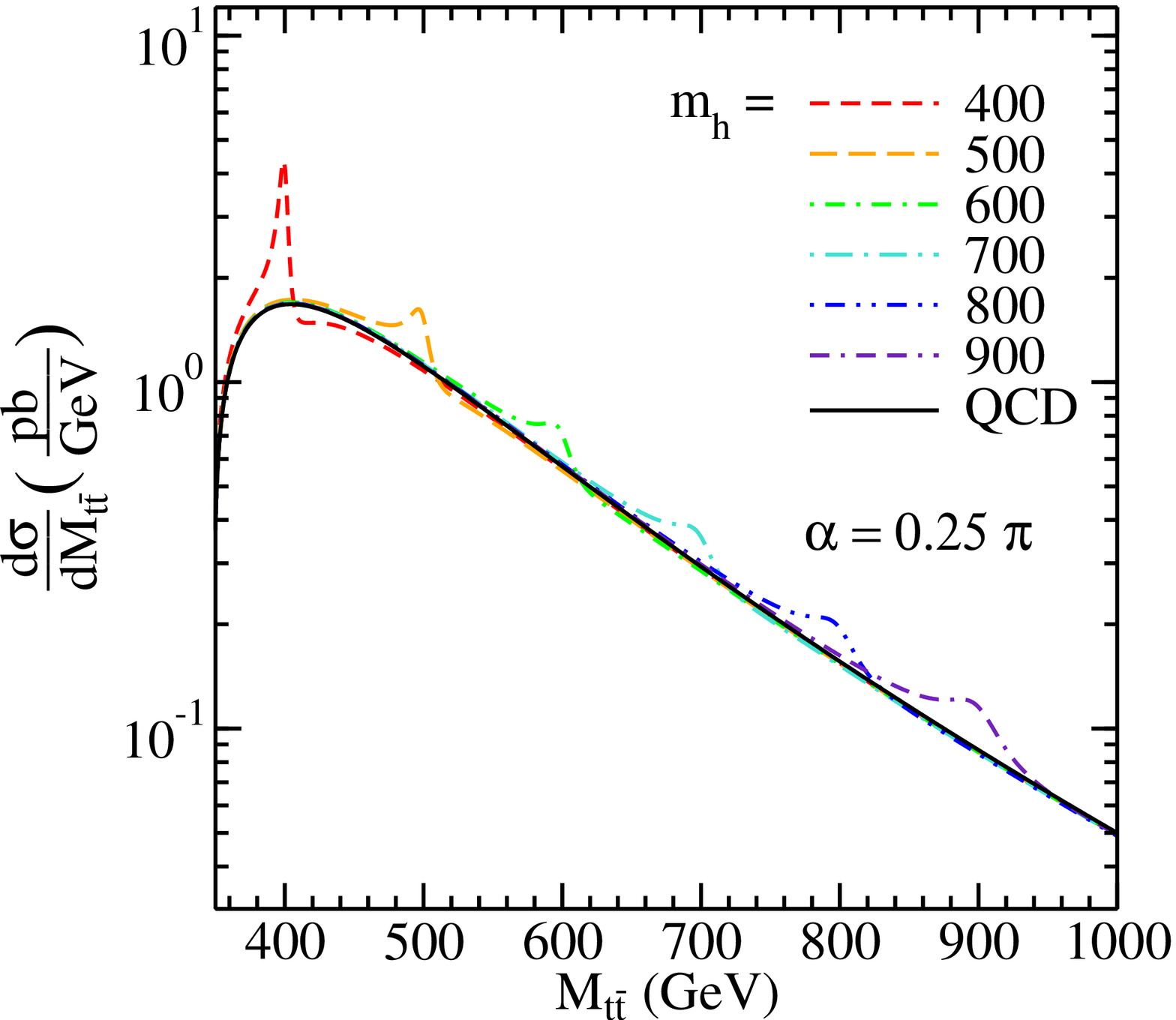}} \\
\subfigure[]{\includegraphics[clip,width=1.00\textwidth]{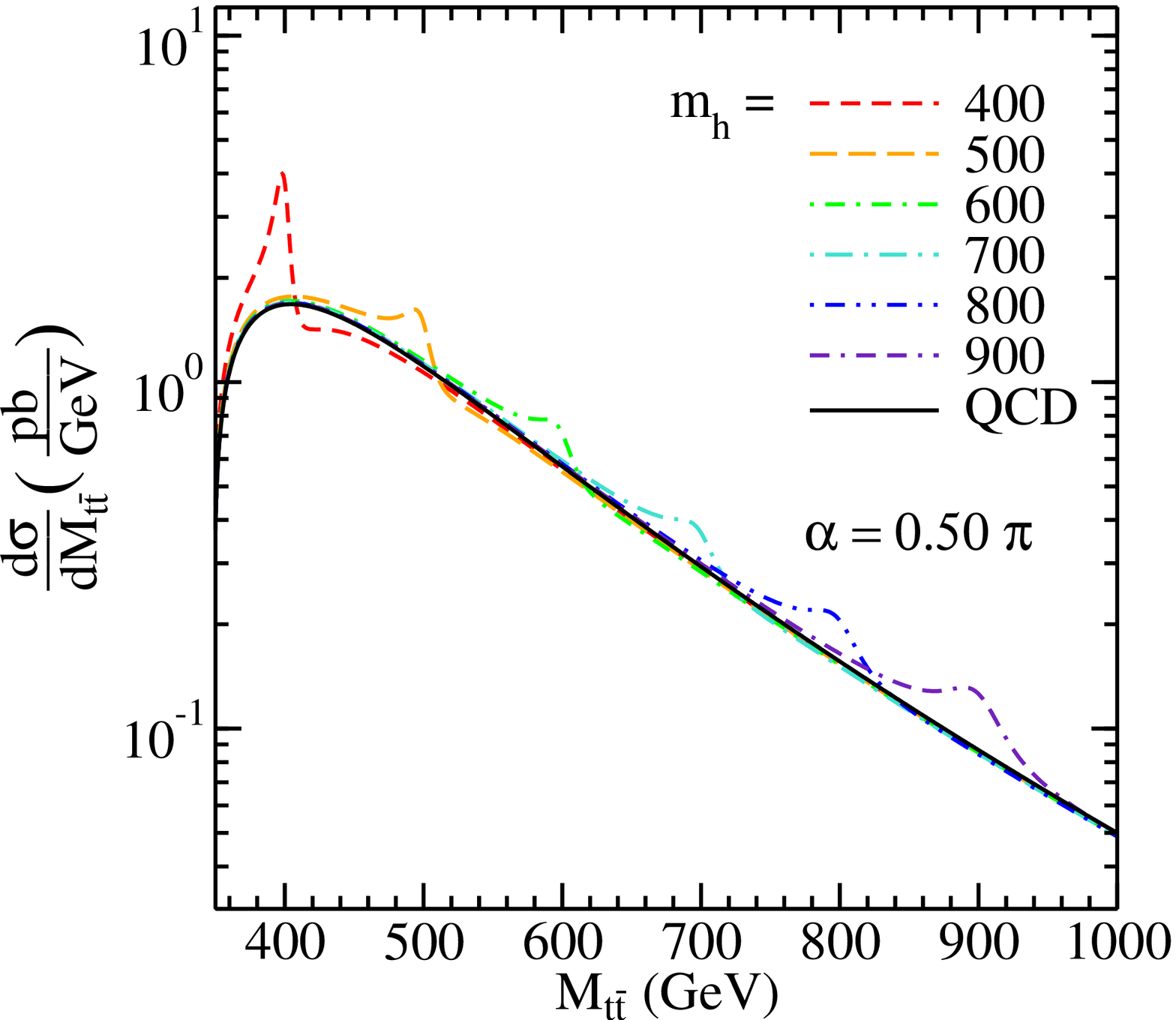}} 
\end{minipage}
\caption{Invariant mass distributions of the top pair from the process $gg \to t \tbar$ at the LHC with $\ecm = 14$~TeV for the top-only (a)-(c) and fourth-generation (d)-(f) scenarios.  These are calculated for several different values of scalar masses and for the phase $\alpha$.  The magnitude of the normalized scalar-top coupling $A$ is fixed at $|A|=1$.}
\label{fig:mtt}
\end{figure}

We plot the invariant mass distribution of the top pair from $gg \to t \tbar$ in Figs.~\ref{fig:mtt}(a)-\ref{fig:mtt}(f) for QCD and for the top-only and fourth-generation scenarios we consider.  We use several scalar masses ranging from $m_{h}=400$~GeV to $900$~GeV and the phases $(\alpha / \pi)=0, 0.25,$ and $0.50$.  In calculating these distributions, we have used a total center-of-mass energy $\ecm = 14$~TeV, CTEQ6L1 PDFs for proton-proton collisions \cite{Pumplin:2002vw}, and fixed renormalization and factorization scales of $\mu_{R} = \mu_{F} = 350$~GeV$\approx 2 m_{t}$.  In addition, we have used one-loop running of $\alpha_{s}$ evaluated at $\mu_{R}$.  These results are in good agreement with those of Refs.~\cite{Dicus:1994bm,Bernreuther:1997gs,Frederix:2007gi}.

When comparing the two scenarios for a fixed value $|A|=1$, it is clear that for a given scalar mass and phase $\alpha$, the fourth-generation scenario gives a much larger deviation from the QCD prediction.  This is due to the well-known enhancement to the scalar production via gluon fusion resulting from the existence of new heavy quarks running in the quark loops of Fig.~\ref{fig:diagrams_h} \cite{Gunion:1989we}.  Indeed, while the fourth-generation scenario provides a scalar signal that appears observable in principle up to the largest mass considered here ($m_{h}=900$~GeV), the top-only case does not have easily observable signals much above the top quark production threshold to around $m_{h}\approx 500$~GeV.  This would not be much improved for larger values of $|A|$, as this tends to simply broaden the distributions seen in Figs.~\ref{fig:mtt}(a)-\ref{fig:mtt}(f).

For a given scenario and given scalar mass, the effect of a variation in $\alpha$ on the top pair invariant mass distribution is twofold:  as seen in Eqs.~(\ref{eq:sigma_h}) and (\ref{eq:sigma_int}), there is a larger threshold suppression of both the signal and interference contributions to the total cross section as $\alpha \to 0$ (the case of a $CP$-even scalar), implying smaller overall signals relative to the case for when $\alpha \to 0.5 \pi$.  However, there is a corresponding suppression in the scalar decay width of Eq.~(\ref{eq:hwidth}).  For smaller scalar masses, the result is larger, narrower peaks in the invariant mass distribution as $\alpha \to 0$, seen most dramatically in the case of $m_{h}=400$~GeV.  At higher scalar masses, however, larger peaks are seen in the limit $\alpha \to \pm 0.5 \pi$.

\begin{figure}
\centering
\includegraphics[clip,width=0.80\textwidth]{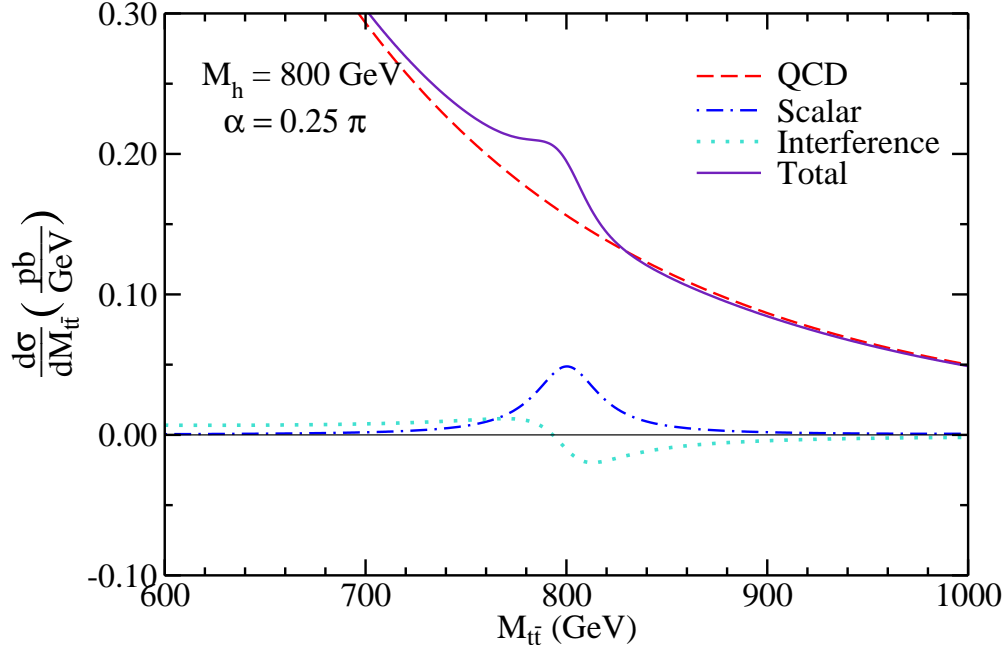} 
\caption{Invariant mass distributions of the top pair from the process $gg \to t \tbar$ at the LHC with $\ecm = 14$~TeV for the fourth-generation scenarios with a scalar mass $m_{h}=800$~GeV and phase $\alpha=0.25\pi$, showing the individual signal, QCD, and interference contributions to the overall distributions.  The magnitude of the normalized scalar-top coupling $A$ is fixed at $|A|=1$.}
\label{fig:mtt_comp}
\end{figure}

As an example showing the direct relationship between the different contributions to the invariant mass distribution, we plot the signal, interference, and QCD terms for a scalar mass $m_{h}=800$~GeV and $\alpha=0.25\pi$ in the fourth-generation scenario in Fig.~\ref{fig:mtt_comp}.  As discussed above, the interference term is non-negligible when compared to the signal and switches sign for a value of $\rts=M_{t\tbar}$ that is somewhat less than the location of the signal resonance at $\rts=m_{h}$.  This results in a peak of the total distribution that is shifted from the resonance, followed by a modest, extended dip relative to the purely QCD case.  The negative values taken by the interference term are important to note, as they serve not only to reduce the total cross section in these regions but will also result in a phase shift of $\pi$ in the interference contribution to the normalized azimuthal distributions, a point we will return to below.

\section{Azimuthal Correlations} \label{sec:correlations}

Following the same procedure described in Sec.~\ref{sec:qcd} and making use of the same event geometry, we calculate the normalized, partonic distributions for the angle $\Phi=(\phi_{2}-\phi_{1})$ for the signal and interference terms separately.  These are given by

\be
\label{eq:dist_h_long}
\frac{1}{\hat{\sigma}_{\rm H}} \frac{d \hat{\sigma}_{\rm H}}{d \Phi}  =  
\left(\frac{1}{2\pi}\right)
\left[1 -  
\left(\frac{\pi}{4}\right)^{2}
\frac{\left(1 - 2 \rho_{w} \right)^{2}}{\left(1 + 2 \rho_{w} \right)^{2}}
\frac{(A_{R}^{2} \beta_{t}^{2} - A_{I}^{2})\cos\Phi + (2 A_{I}A_{R} \beta_{t}) \sin\Phi}{(A_{R}^{2} \beta_{t}^{2} + A_{I}^{2})}
\right]
\ee

\noi and

\bea
\label{eq:dist_int_long}
\frac{1}{\hat{\sigma}_{\rm INT}} \frac{d \hat{\sigma}_{\rm INT}}{d \Phi}  =  
\left(\frac{1}{2\pi}\right)
\Bigg[1 -  
\left(\frac{\pi}{4}\right)^{2} 
\frac{\left(1 - 2 \rho_{w} \right)^{2}}{\left(1 + 2 \rho_{w} \right)^{2}}
& \Bigg\{ &
\frac{(2 A_{R}^{2} \beta_{t}^{2} C_{H} - 3 A_{I}^{2} C_{A})}{(2 A_{R}^{2} \beta_{t}^{2} C_{H}+ 3A_{I}^{2} C_{A})}
\cos\Phi  \\ \nn
& + & 
\frac{(A_{I} A_{R} \beta_{t})(3 C_{A} + 2 C_{H})}{(2 A_{R}^{2} \beta_{t}^{2} C_{H}+ 3A_{I}^{2} C_{A})}
\sin\Phi
\Bigg\}
\Bigg] \, ,
\eea

\noi where $C_{H,A}$ and $\rho_{w}$ are as defined above.   These can be greatly simplified and more easily interpreted if we now define phases $\chi$ and $\xi$ according to

\bea
\label{eq:phases_def}
A_{\chi}  = & \left(A_{R} \beta_{t} + i A_{I} \right)  & =  |A_{\chi}| e^{i \chi} \\ \nn
A_{\xi}    = & \left[ (2 A_{R} \beta_{t} C_{H}) +  i (3 A_{I} C_{A}) \right] & =  |A_{\xi}| e^{i \xi} \, .
\eea

\noi These are related to the phase $\alpha$, the main parameter of interest, by

\bea
\label{eq:phase_relations}
\tan\chi & = & \frac{A_{I}}{\beta_{t} A_{R}} = \left(\frac{1}{\beta_{t}}\right) \tan{\alpha} \\ \nn
\tan\xi & = & \frac{3 A_{I} C_{A}}{2 A_{R} \beta_{t} C_{H}}  =  \left( \frac{\textrm{3 Re}\left(F_{A} \prophp \right)}{\textrm{2 Re}\left(F_{H} \prophp \right)} \right) \left(\frac{1}{\beta_{t}}\right)  \tan{\alpha} \, .
\eea

\noi With the above definitions, Eq.~(\ref{eq:dist_h_long}) and (\ref{eq:dist_int_long} become

\be
\label{eq:phi_h}
\frac{1}{\hat{\sigma}_{\rm H}} \frac{d \hat{\sigma}_{\rm H}}{d \Phi} = 
\left(\frac{1}{2\pi}\right)
\left[ 1 - 
\left(\frac{\pi}{4}\right)^{2} 
\frac{\left(1 - 2 \rho_{w} \right)^{2}}{\left(1 + 2 \rho_{w} \right)^{2}}
\cos(\Phi - 2 \chi) 
\right]
\ee

\noi and

\be
\label{eq:phi_int}
\frac{1}{\hat{\sigma}_{\rm INT}} \frac{d \hat{\sigma}_{\rm INT}}{d \Phi} = 
\left(\frac{1}{2\pi}\right)
\left[ 1 - 
\left(\frac{\pi}{4}\right)^{2}
\frac{\left(1 - 2 \rho_{w} \right)^{2}}{\left(1 + 2 \rho_{w} \right)^{2}}
\frac{\cos(\Phi - (\chi + \xi))}{\cos(\chi - \xi)} 
\right] \, .
\ee

\begin{figure}
\centering
\includegraphics[clip,width=0.80\textwidth]{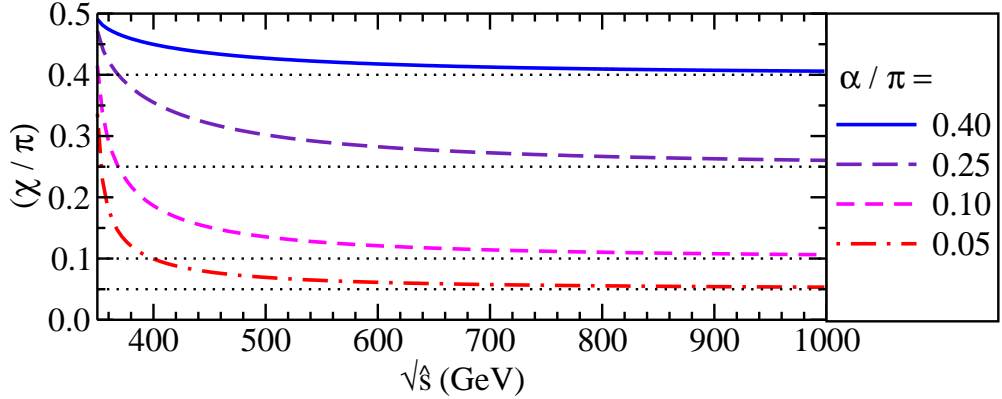}
\caption{The value of the phase $\chi$ as a function of $\rts$ for several different values of the phase $\alpha$.  The dotted lines give the corresponding fixed values of $\alpha$.  For $\alpha<0$, the resulting behavior is the mirror image about the $\rts$ axis.}
\label{fig:chi}
\end{figure}

\noi As discussed in Refs.~\cite{Barger:2011tb,Baumgart:2011wk}, the phase $\chi$ appearing in Eq.~(\ref{eq:phi_h}) gives a rather direct experimental probe of the $CP$ nature of the scalar particle and the phase $\alpha$ when considering the signal distribution in isolation: for highly boosted tops, and therefore large values of $\beta_{t}$, $\chi \approx \alpha$ and there is a phase shift in the $\Phi$ distribution of $\approx - 2 \alpha$.  The precise behavior of $\chi$ as a function of $\rts$ is given for several different values of $\alpha$ in Fig.~\ref{fig:chi}.  We note that in the $CP$-even ($\alpha=0$) and $CP$-odd ($\alpha=0.5\pi$) cases, $\chi$ takes the same value as $\alpha$ independent of $\beta_{t}$.  In the $CP$-odd case, there is then a phase shift of $\pi$ and the signal distribution is exactly in phase with the QCD-only distribution, while in the $CP$-even case, the two distributions are correspondingly out of phase by $\pi$.

What is more interesting, however, is the possibility to detect $CP$ violation by means of a small phase shift in this distribution, a quantity which is often measured using tools such as "up-down" (or similar) asymmetries \cite{Chang:1993jy,Bernreuther:1998qv}, essentially comparing the values for $\Phi>0$ to those with $\Phi<0$.  While the full $\Phi$ distribution is perhaps more difficult to measure than an asymmetry -- if only for purely statistical reasons -- it is interesting in that it contains more information: not only does it distinguish between the $CP$-even and $CP$-odd cases, which an up-down asymmetry does not, but it shows how simply and directly the violation is related to the phase $\alpha$.

It is also interesting to note that the normalized signal distribution at the parton level is completely characterized by the one parameter $\alpha$ through the phase $\chi$; it carries no information of the magnitude $|A|$, the total width $\Gamma_{h}$, nor of any other details concerning the production of the scalar itself.  This is due to the scalar nature of the signal:  because there are no spin correlations in the production and decay of $h$, the normalized $\Phi$ distribution for $\sigma(gg \to h \to t \tbar \to b \bbar \wplus \wminus)$ is the same as that for the partial width $\Gamma(h \to t \tbar \to b \bbar \wplus \wminus)$.  Normalized thus, Eq.~(\ref{eq:phi_h}) is a fairly generic result for scalars which couple and decay to weakly decaying particles via a coupling like Eq.~(\ref{eq:htt}), a point noted in \cite{Barger:2011tb}.

\begin{figure}
\centering
\subfigure[]{\includegraphics[clip,width=0.80\textwidth]{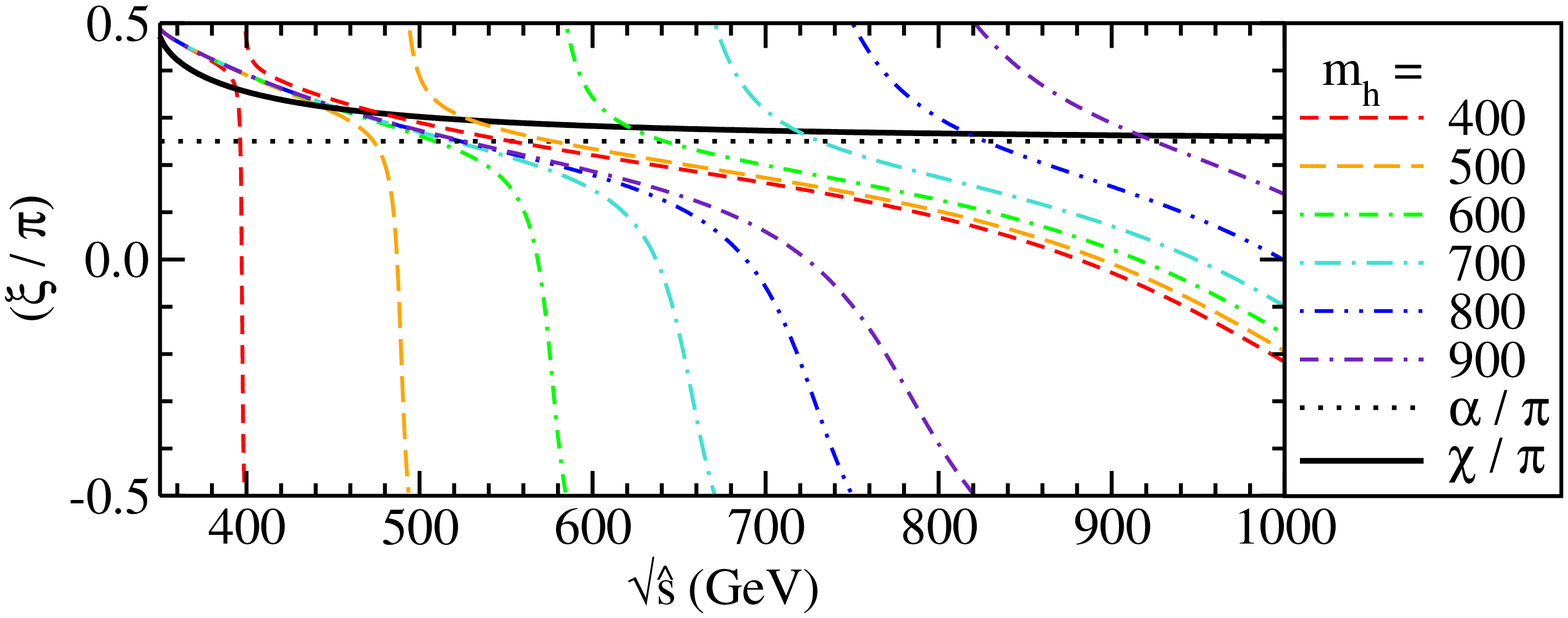}} \\
\subfigure[]{\includegraphics[clip,width=0.80\textwidth]{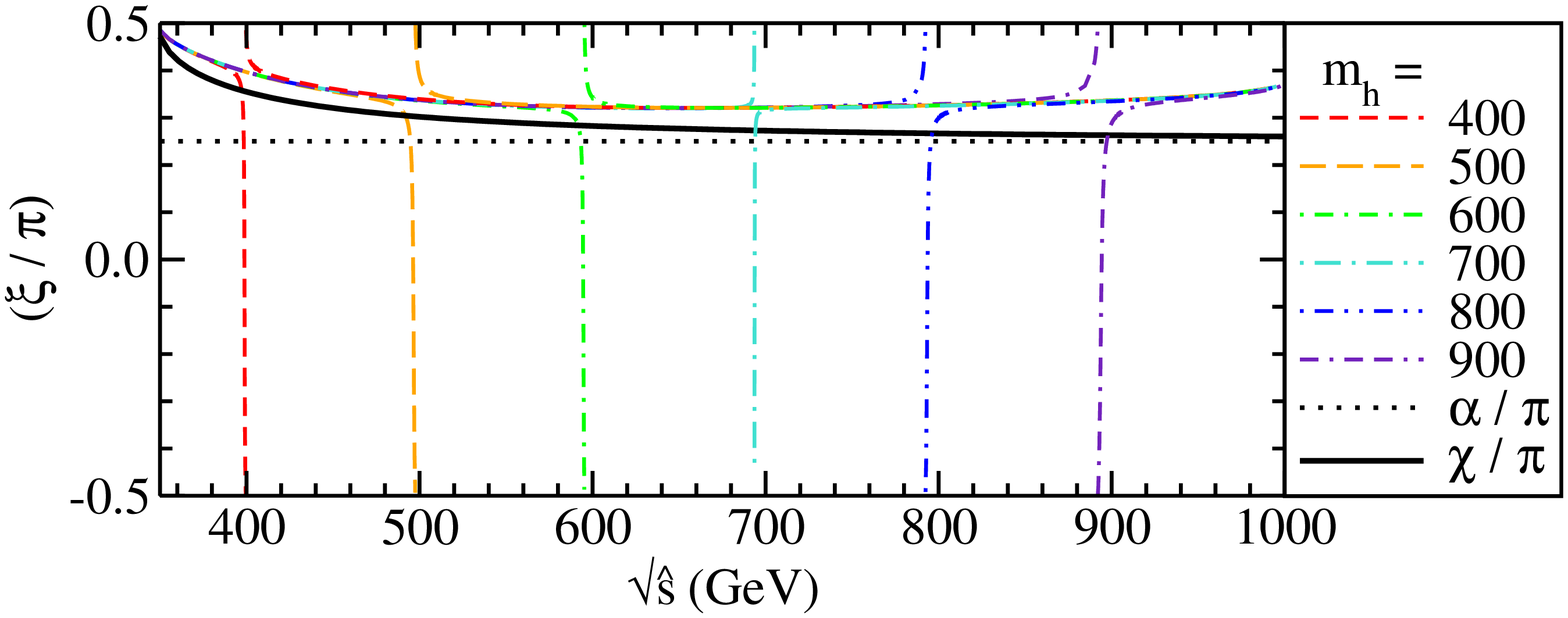}} 
\caption{The value of the phase $\xi$ as a function of $\rts$ for the choice $\alpha=0.25\pi$ and several values of the scalar mass $m_{h}$ for the (a) top-only scenario and the (b) fourth-generation scenario.  This is compared to the value of $\chi$, also as a function of $\rts$, plotted on the same scale and the fixed value of $\alpha=0.25\pi$.}
\label{fig:phases}
\end{figure}

While the normalized distribution of the signal can be completely characterized by the phase $\chi$, the distribution derived from its interference with the QCD process also depends on the phase $\xi$.  Like $\chi$, the phase $\xi$ takes on the same value as $\alpha$ for purely $CP$-even or $CP$-odd couplings.  In these cases -- and more generally for cases in which $\xi=\chi$ -- the normalized distribution derived from the interference term takes the exact same form as that for the signal and will therefore serve only to enhance or minimize the effect of the signal depending on the sign of $\hat{\sigma}_{\rm INT}$.  However, in the $CP$-violating case the behavior of $\xi$, and therefore the effect of the interference on the  $\Phi$ distribution, is more complicated.  We plot the value of $\xi$ as a function of $\rts$ in Figs.~\ref{fig:phases}(a) and \ref{fig:phases}(b) for the fixed value $\alpha=0.25\pi$ and for several different values of the scalar mass $m_{h}$.  This is done for both the top-only and the fourth-generation scenarios.  In the context of its use in Eq.~(\ref{eq:phi_int}), we can restrict $\xi$ to the range $- \pi/2 < \xi < \pi/2$ without loss of generality.  This will be assumed for the remainder of this paper.

For a given scalar mass, there is a region in both the top-only and fourth-generation cases where $\xi$ switches sign, taking on its full range of values in the process.  This occurs near, but not generally at, $\rts = m_{h}$ and is again related to the pole in the propagator appearing in the functions $C_{H}$ and $C_{A}$.  The complications that would arise from this are generally minimized by the fact that the interference contribution to the partonic cross sections is switching sign here as well and is therefore in the region where it contributes least to the overall behavior of the full distribution.  What is more important, then, is the behavior of $\xi$ outside this region, where the interference plays a much larger role.  It is here that the two cases (top-only and fourth-generation) are quite different.  Outside of the pole region, the value of $\xi$ in the fourth-generation case takes a relatively stable value lying slightly apart from $\chi$ for all scalar masses, while no such behavior is seen for the top-only case.  This is rooted in the different contributions that the top quark and the fourth-generation quarks make to the functions $F_{H}$ and $F_{A}$.  The new heavy quarks have $\tau_{Q} > 1$ such that their contributions are purely real while the top has $\tau_{t}<1$ and therefore gives a complex contribution.  The former contributions dominate in the fourth-generation case, minimizing the effect of the complex part of the scalar propagator, and therefore the scalar width, in $C_{H,A}$.  This results in the ratio $C_{A} / C_{H} \approx 1$ in Eq.~(\ref{eq:phase_relations}) for lower values of $\rts$, which then grows slightly approaching the $\tau_{Q} = 1$ threshold at $\rts=1$~TeV.  The main conclusion that can be drawn here is that in the fourth-generation case, the interference term will give contributions to the overall $\Phi$ distribution that are approximately in phase with the pure signal contribution as long as $\hat{\sigma}_{\rm INT}(\shat) >0$, while in the top-only case this is much more dependent on the scalar mass and the particular top pair invariant mass region of interest.

Now, given all the individual contributions to the $\Phi$ distribution, the complete observable behavior will be given by

\be
\frac{1}{\sigma_{tot}} \frac{d \sigma_{tot}}{d \Phi} = \frac{1}{\sigma_{tot}} \left(
\frac{d \sigma_{\rm QCD}}{d \Phi} +
\frac{d \sigma_{\rm H}}{d \Phi} +
\frac{d \sigma_{\rm INT}}{d \Phi}
\right) \, ,
\ee

\noi where each $\sigma_{i}$ signifies that the parton-level results have been integrated with PDFs over some region of top-pair invariant mass.  We note here that for each contribution we can define the normalized distribution and relative weight as

\bea
f_{i}(\Phi) &  \equiv & \frac{1}{\sigma}_{i} \frac{d \sigma_{i}}{d \Phi}  \\
w_{i} & = & \left( \frac{\sigma_{i}}{\sigma_{tot}} \right)
\eea

\noi such that the total distribution can be given as

\be
f_{tot}(\Phi) = \left( w_{\rm QCD} f_{\rm QCD}(\Phi) + w_{\rm H} f_{\rm H}(\Phi) + w_{\rm INT} f_{\rm INT}(\Phi) \right) \, ,
\ee

\noi with $\sum_{i} w_{i} = 1$.  The total $\Phi$ distribution is therefore the sum of the individual normalized contributions weighted by their contribution to the cross section\footnote{We note one slight subtlety here: it can be shown that while $w_{\rm INT}\to0$ as $(\chi - \xi) \to \pm \pi/2$, there will generally be a small non-zero contribution to $f_{tot}(\Phi)$ from the interference term here due to the factor $\cos^{-1}(\chi-\xi)$ appearing in Eq.~(\ref{eq:phi_int}).}.  Like the individual terms, we can write

\be
f_{tot}(\Phi) = \left(\frac{1}{2\pi}\right)
\left(1+ C_{tot}
\left(\frac{\pi}{4}\right)^{2}
\frac{\left(1 - 2 \rho_{w} \right)^{2}}{\left(1 + 2 \rho_{w} \right)^{2}} 
\cos(\Phi +\delta_{tot})
\right) \, ,
\ee

\noi where $C_{tot}$ and $\delta_{tot}$ depend on the values of the various parton-level angular distributions integrated with the PDFs.  We note, however, that when integrated over a small enough range of $\rts=M_{t\tbar}$, these distributions should be a good estimate of the resulting behavior, as the PDF integrations cancel out if the weights and angular distributions are assumed to have little variation in that region.  We can then make the statement that what we are ultimately interested in, then, is if deviations can be seen in one or both of $C_{tot}=C_{QCD}$ and $\delta_{tot} = 0$ due to the effect of a scalar particle coupling to the top quark as described.  

\begin{figure}
\centering
\begin{minipage}{0.45\textwidth}
\subfigure[]{\includegraphics[clip,width=1.00\textwidth]{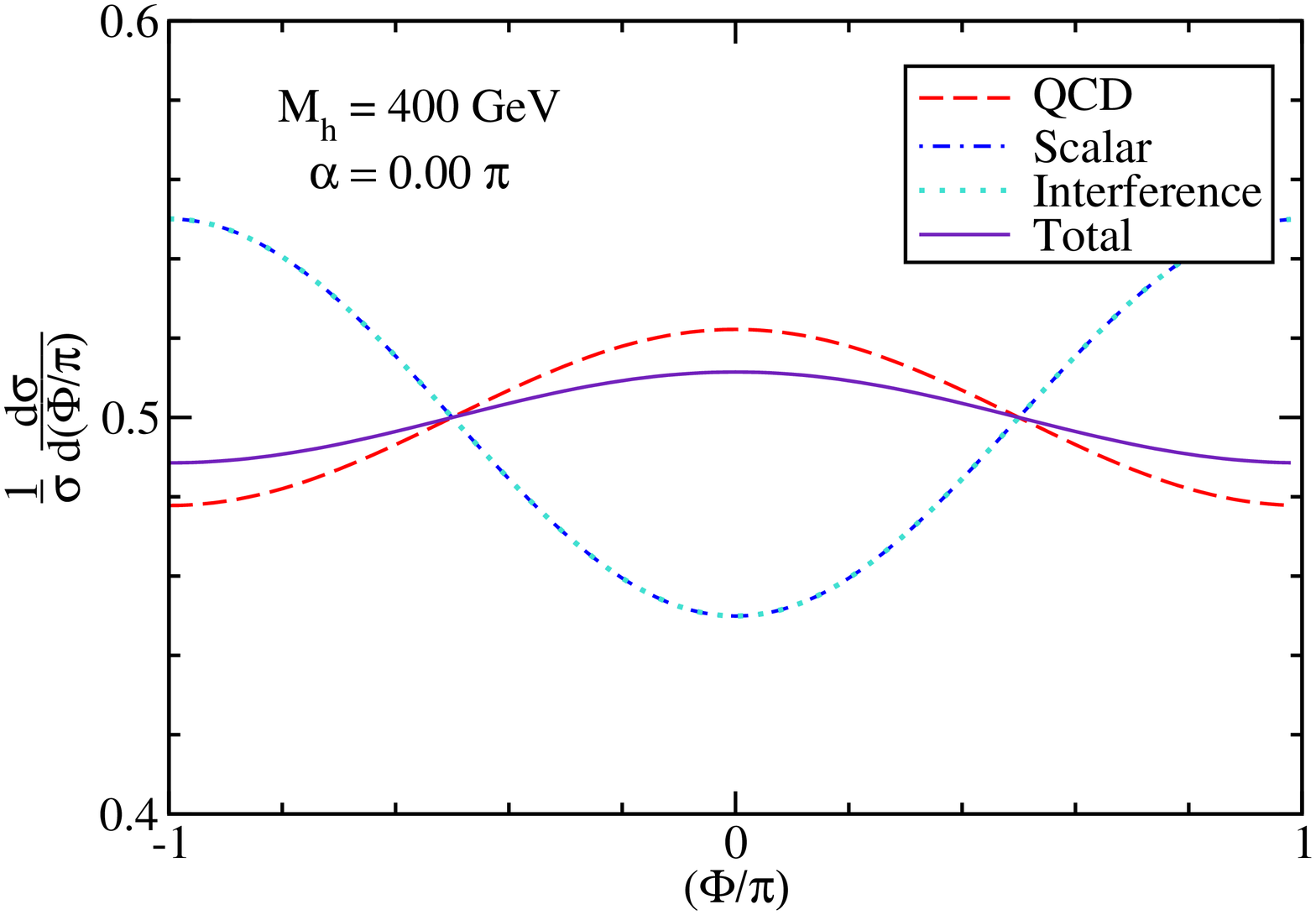}} \\
\subfigure[]{\includegraphics[clip,width=1.00\textwidth]{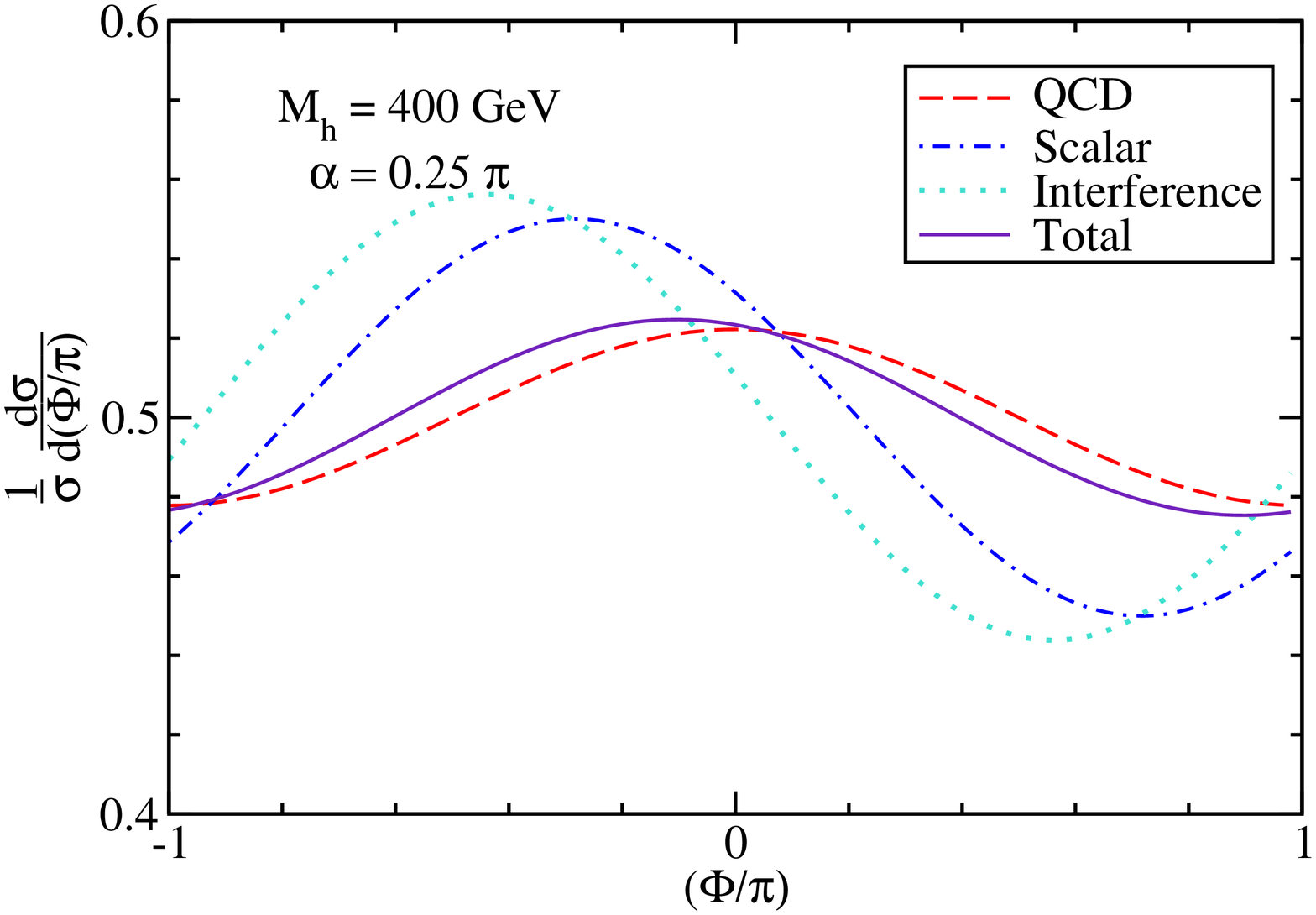}} \\
\subfigure[]{\includegraphics[clip,width=1.00\textwidth]{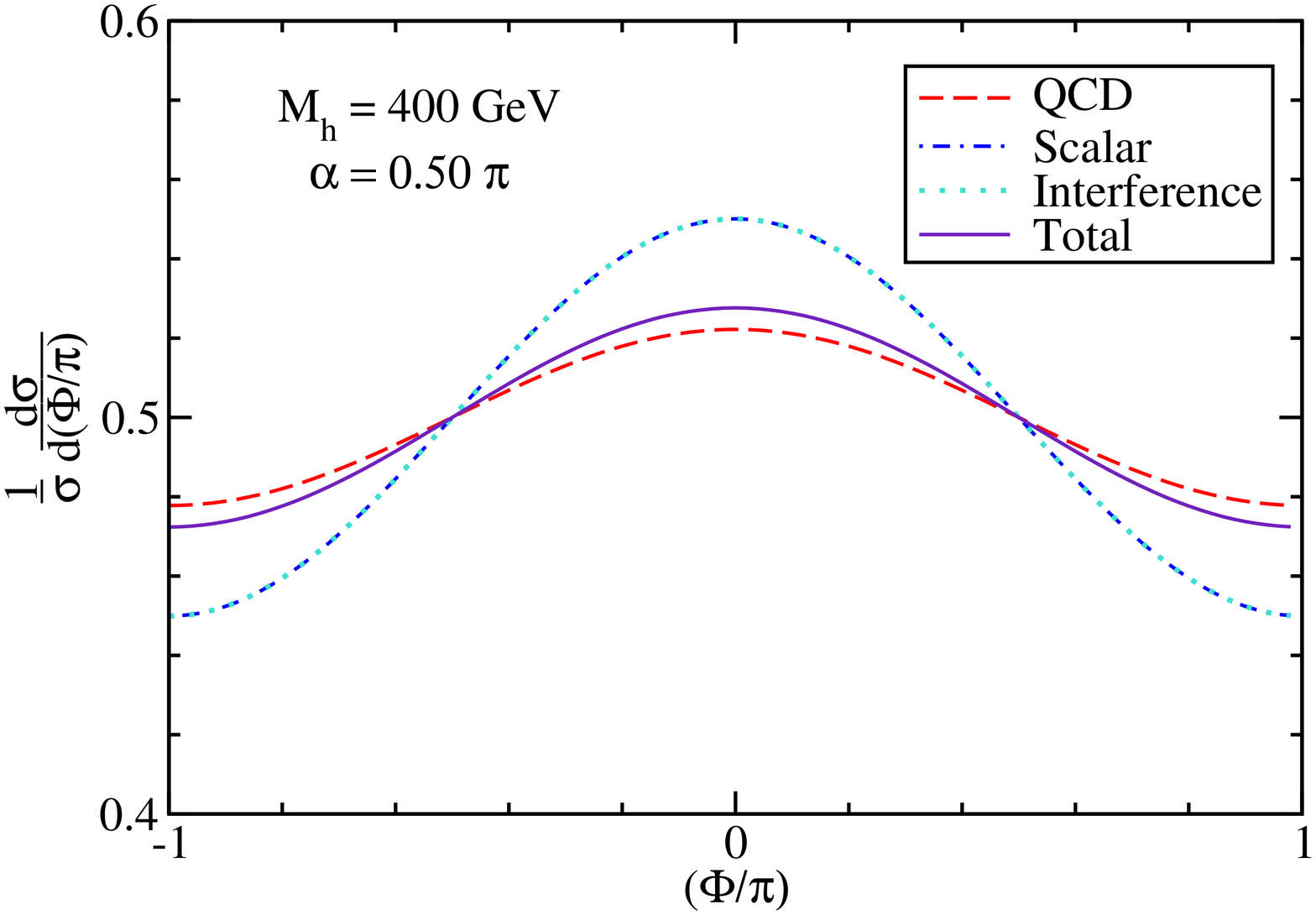}}
\end{minipage} \hspace{1cm}
\begin{minipage}{0.45\textwidth}
\subfigure[]{\includegraphics[clip,width=1.00\textwidth]{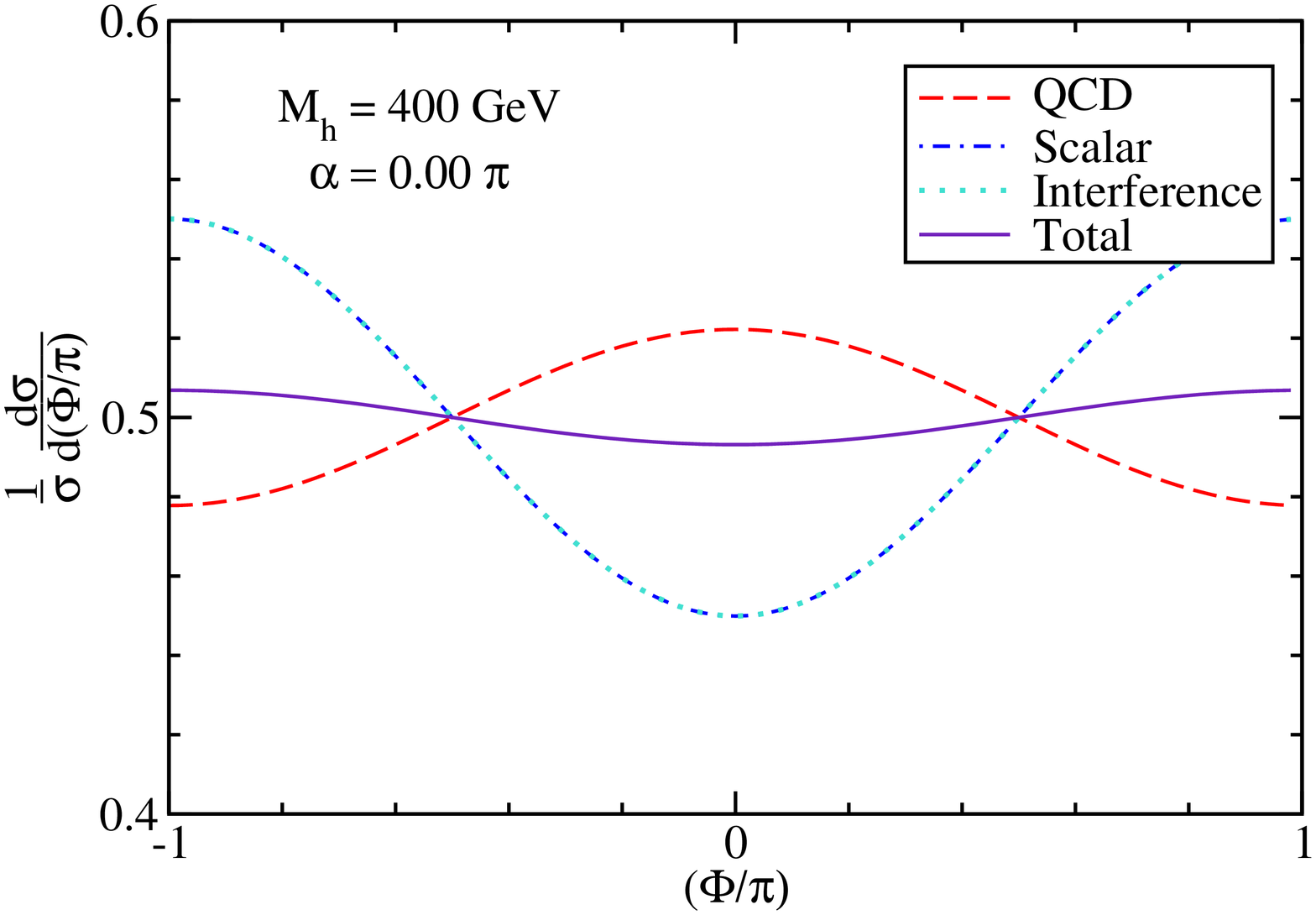}} \\
\subfigure[]{\includegraphics[clip,width=1.00\textwidth]{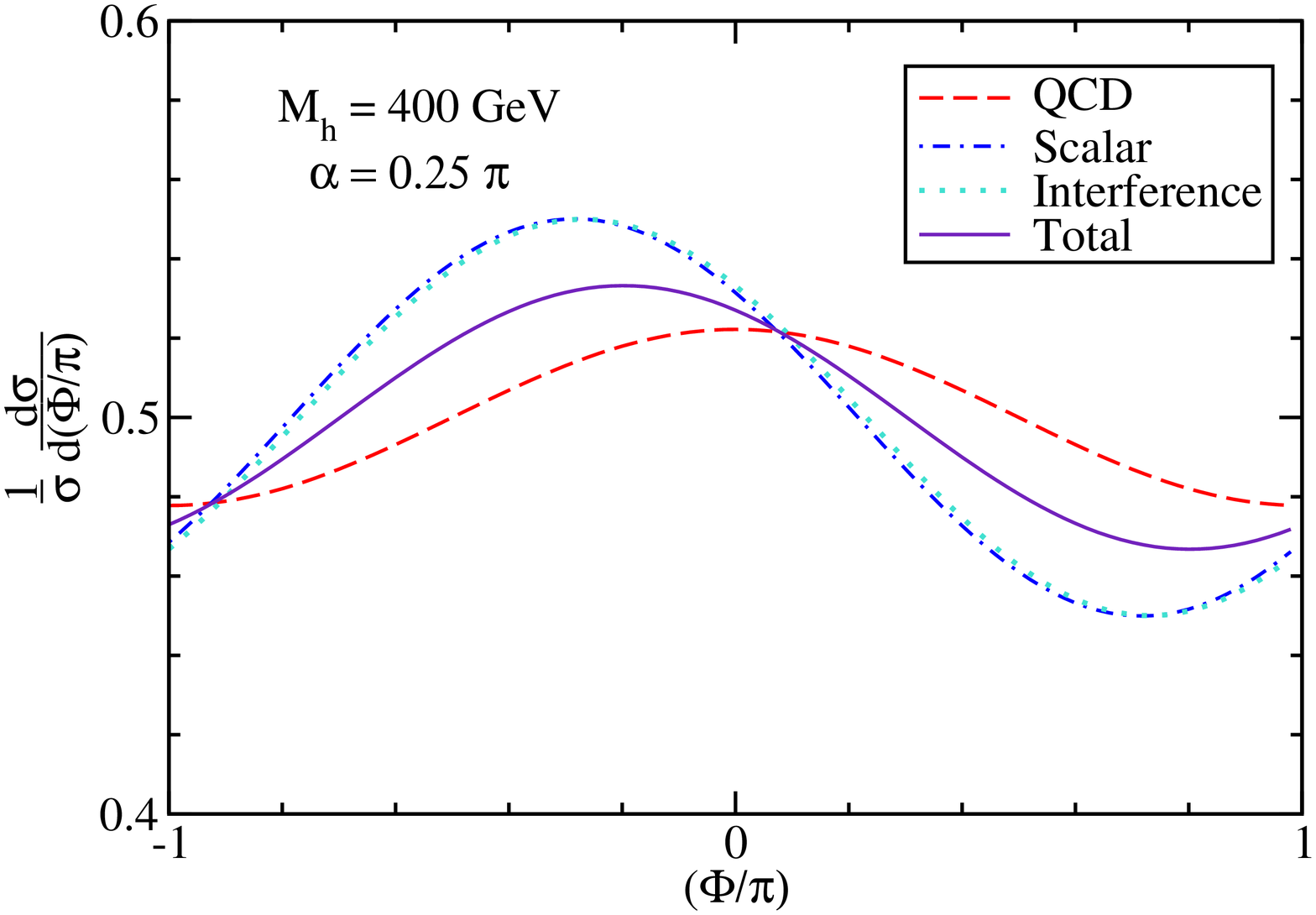}} \\
\subfigure[]{\includegraphics[clip,width=1.00\textwidth]{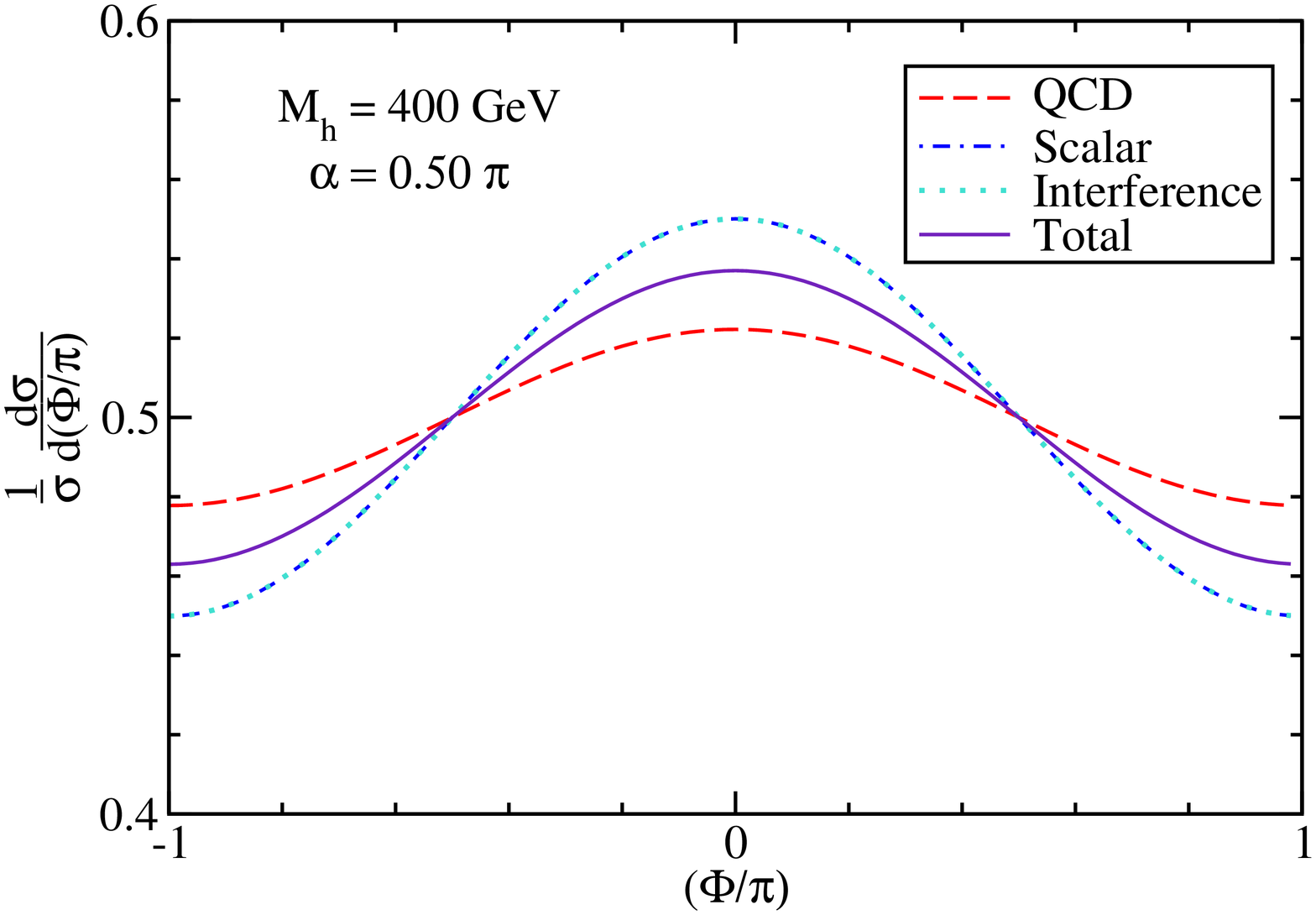}}
\end{minipage}
\caption{Total normalized $\Phi$ distributions for the process $gg \to t \tbar \to b \bbar \wplus \wminus$, in addition to their individual contributions from QCD, the scalar signal, and their interference, at the LHC with $\ecm = 14$~TeV for the top-only (a)-(c) and fourth-generation (d)-(f) scenarios.  These are calculated for the scalar mass $m_{h}=400$~GeV and for the phase $(\alpha / \pi)=0, 0.25, 0.50$ and have been integrated with PDFs in the range $390\gev<\rts<400\gev$.  The magnitude of the normalized scalar-top coupling $A$ is fixed at $|A|=1$.}
\label{fig:phi_400}
\end{figure}

In Figs.~\ref{fig:phi_400}(a)-\ref{fig:phi_400}(f), we plot the total normalized $\Phi$ distributions in addition to the contributions from QCD, the scalar signal, and their interference at the LHC at  $\ecm = 14$~TeV for a scalar mass of $m_{h}=400$~GeV and for $(\alpha / \pi)=0, 0.25, 0.50$.  These distributions have been integrated with PDFs over the range $390\gev<\rts<400\gev$.  This particular range is motivated by the desire to avoid regions above the resonance in which the interference gives a negative contribution, thereby weakening the overall strength of the deviation.  These plots are meant to give an example of the maximum effect the scalar can have on the $\Phi$ distribution, as $m_{h}=400$~GeV gives some of the largest signal cross sections possible, as seen in Fig.~\ref{fig:mtt}(a)-\ref{fig:mtt}(f).  Rather large departures from the QCD prediction are seen, in particular, for the fourth-generation scenario.  For instance, in the $CP$-even case ($\alpha=0$), the resulting distribution is completely out of phase with the QCD prediction, albeit with a reduced amplitude.  This scenario is rather unrealistic when considering this particular scalar mass, however.  Such a large spike in the invariant mass distribution of top pairs would likely have been observed in the early running of the LHC, if not at the Fermilab Tevatron, as this is precisely the region with the largest statistics.

\begin{figure}
\centering
\begin{minipage}{0.45\textwidth}
\subfigure[]{\includegraphics[clip,width=1.00\textwidth]{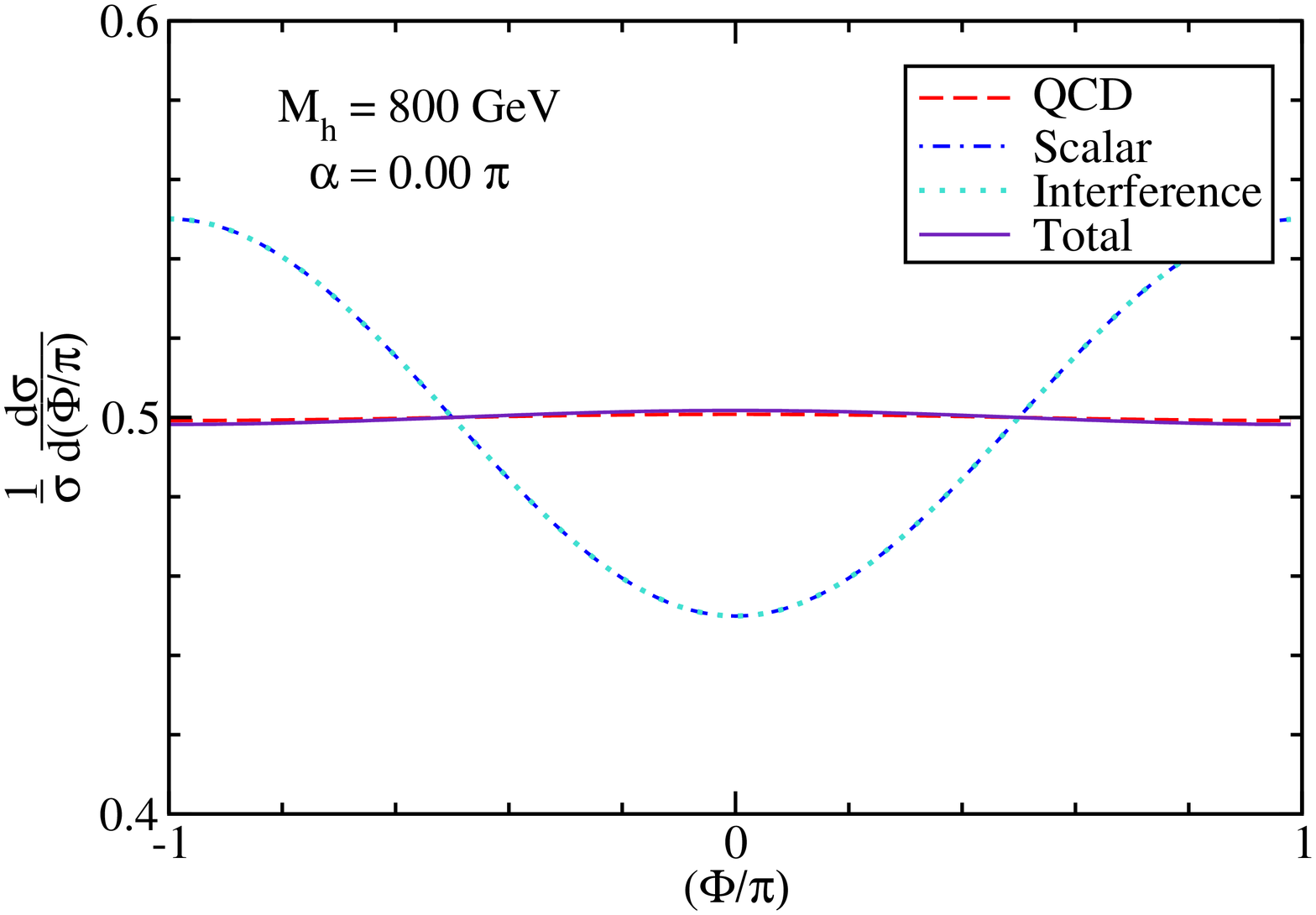}} \\
\subfigure[]{\includegraphics[clip,width=1.00\textwidth]{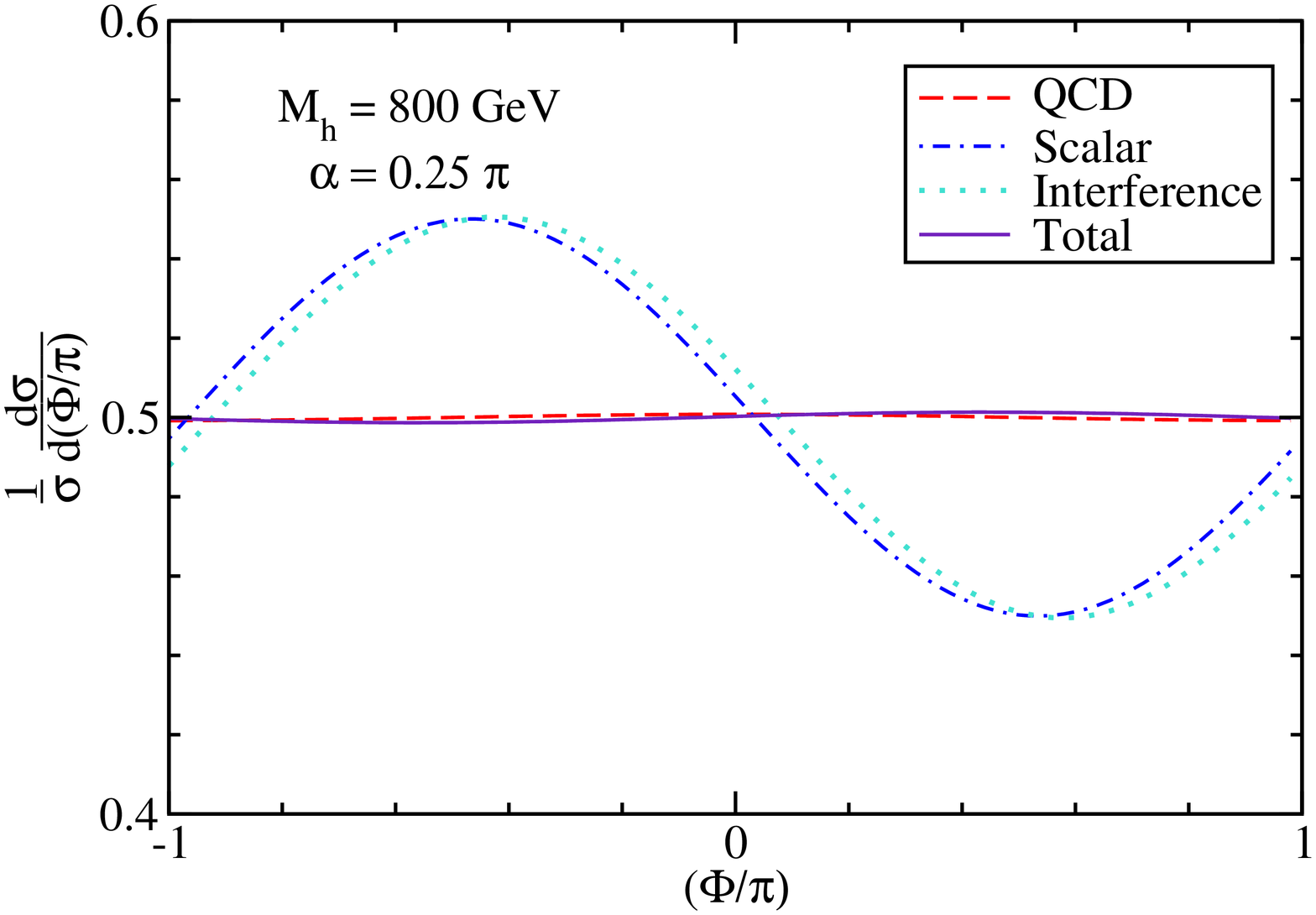}} \\
\subfigure[]{\includegraphics[clip,width=1.00\textwidth]{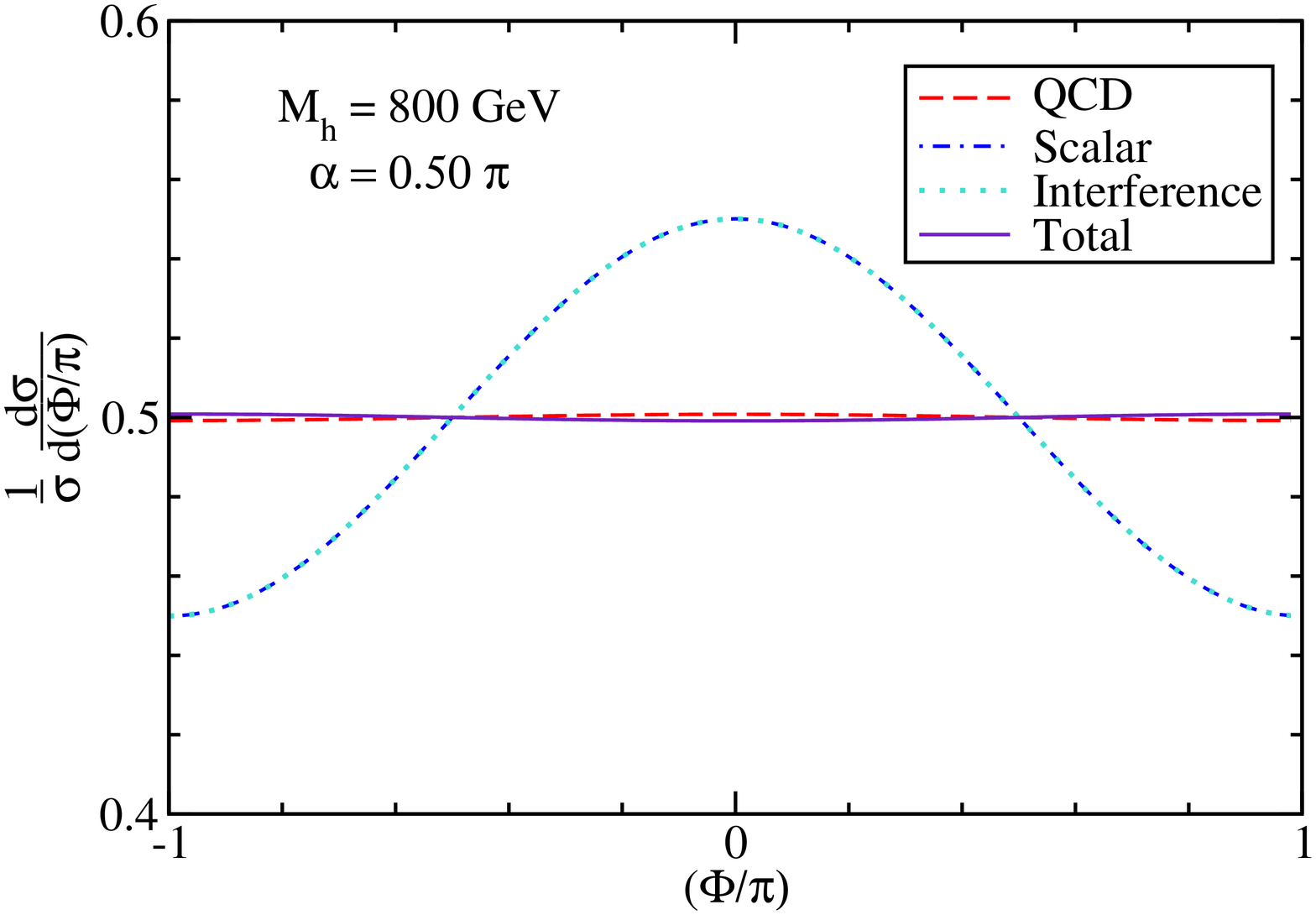}}
\end{minipage} \hspace{1cm}
\begin{minipage}{0.45\textwidth}
\subfigure[]{\includegraphics[clip,width=1.00\textwidth]{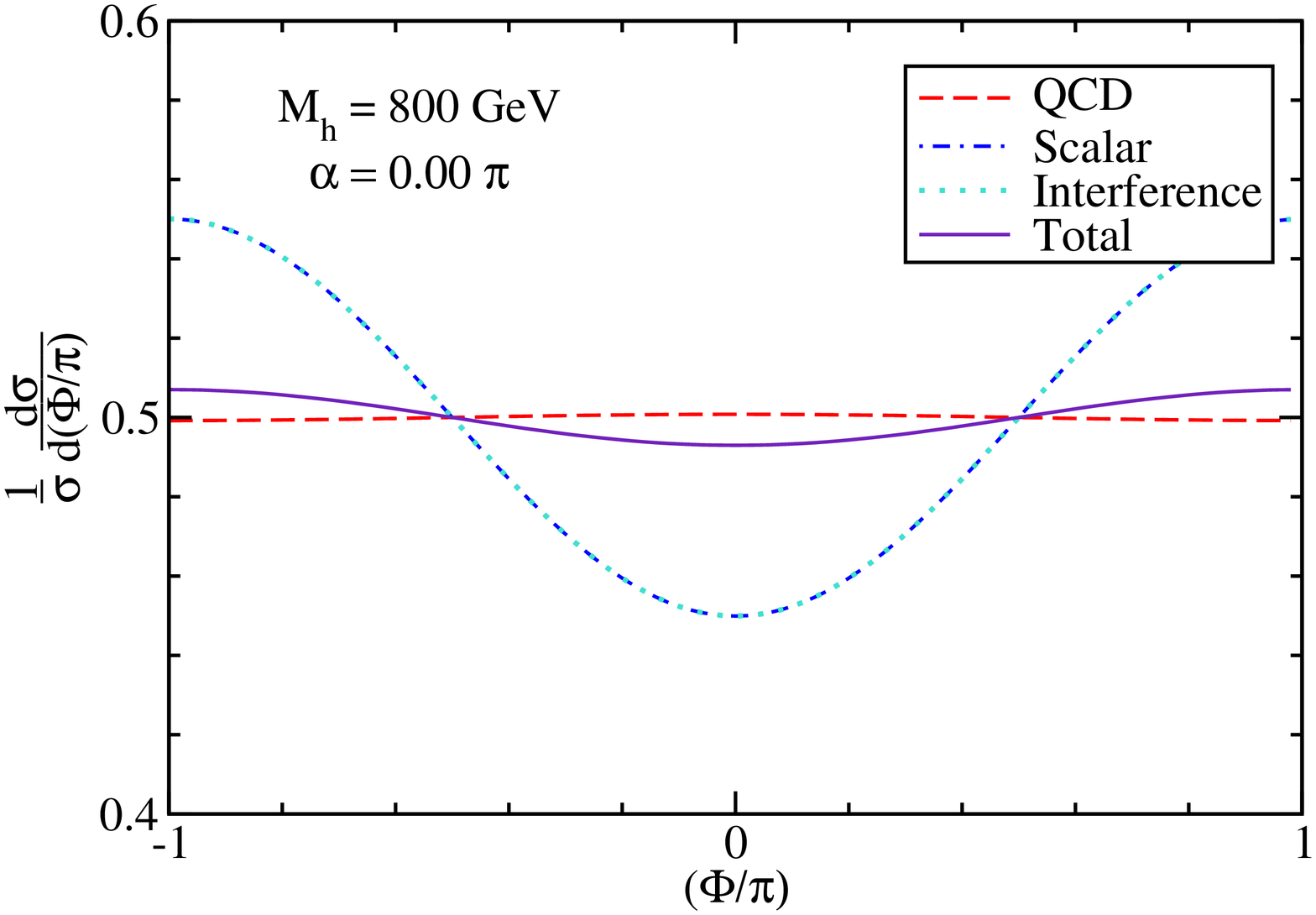}} \\
\subfigure[]{\includegraphics[clip,width=1.00\textwidth]{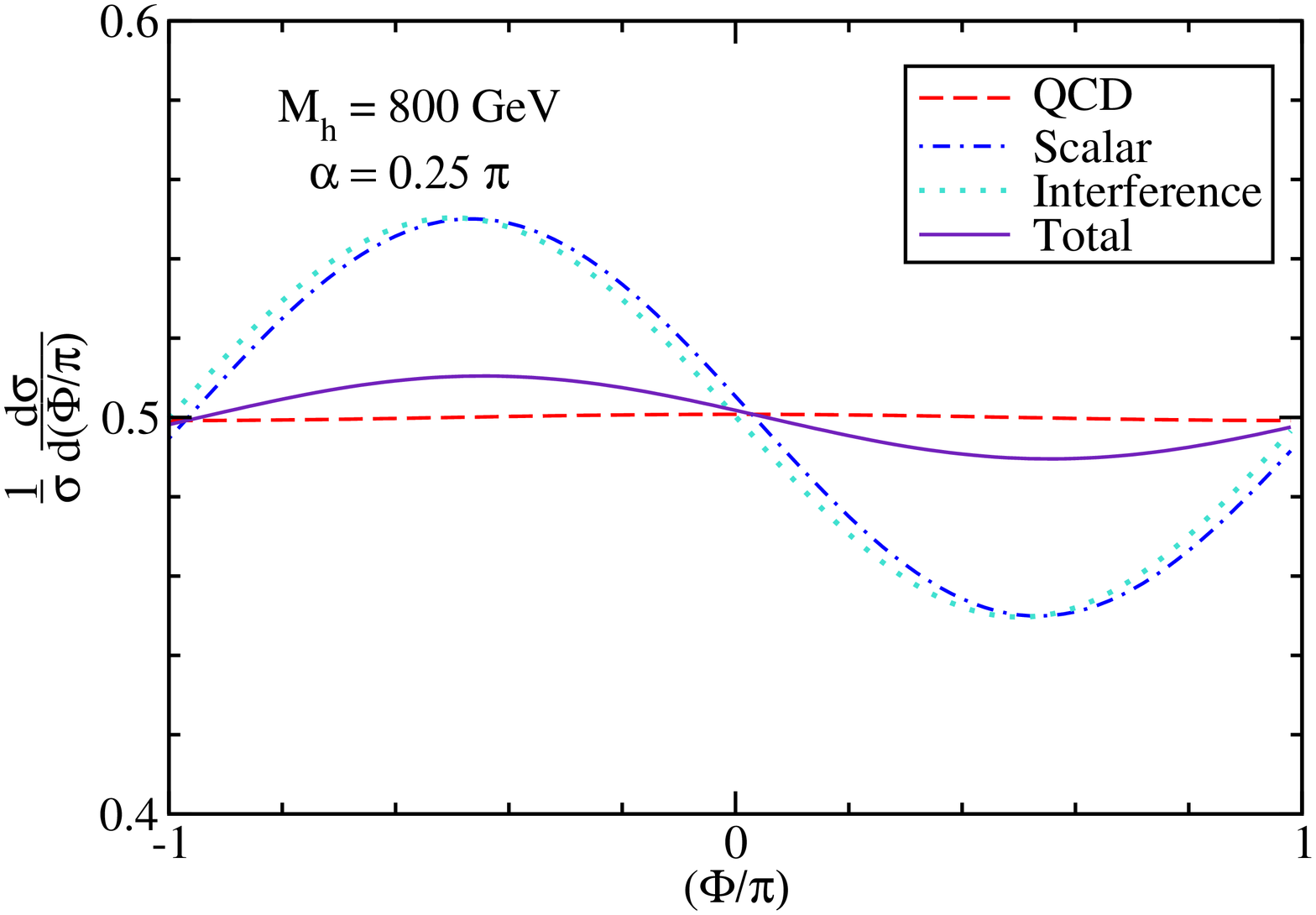}} \\
\subfigure[]{\includegraphics[clip,width=1.00\textwidth]{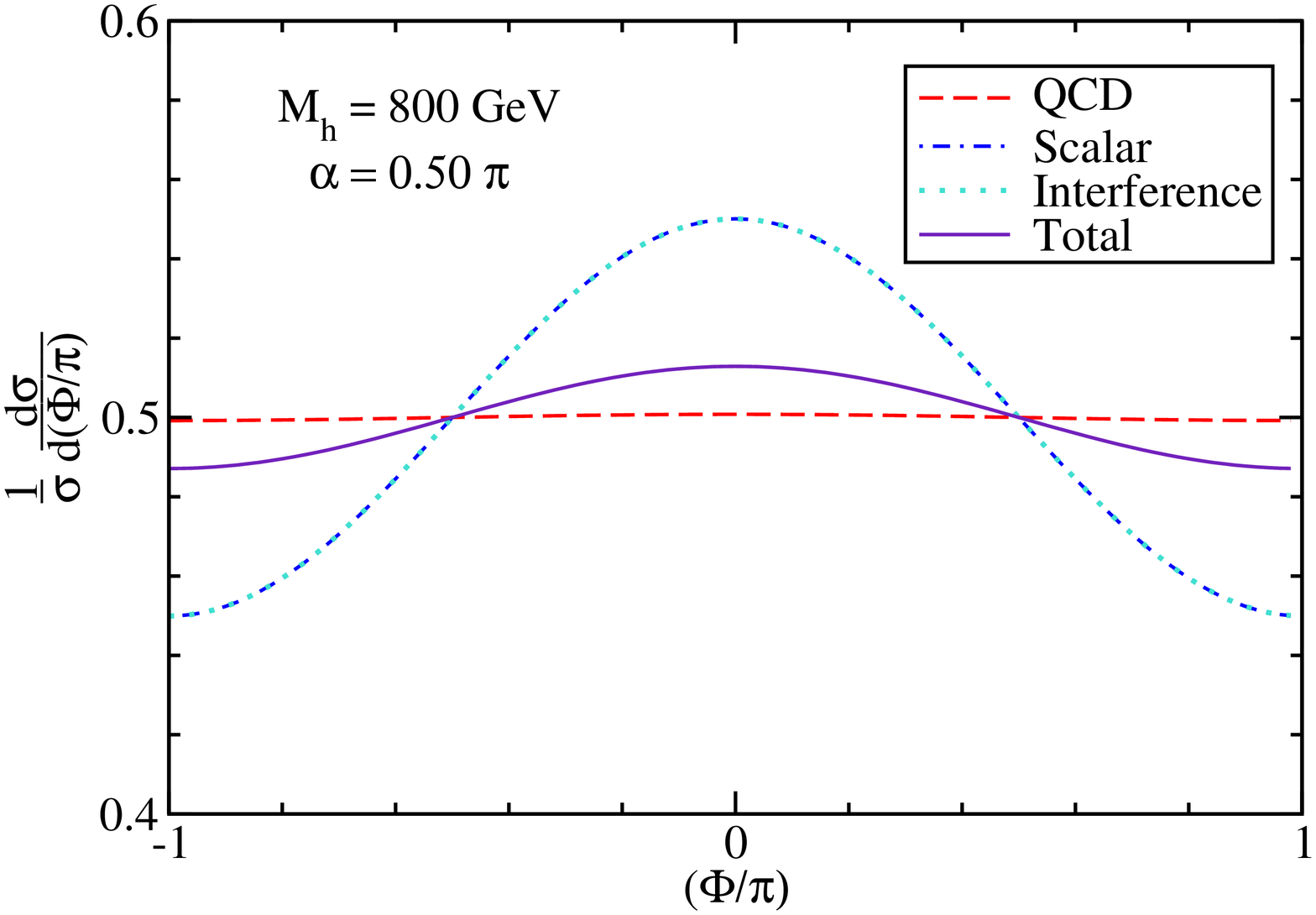}}
\end{minipage}
\caption{Total normalized $\Phi$ distributions for the process $gg \to t \tbar \to b \bbar \wplus \wminus$, in addition to their individual contributions from QCD, the scalar signal, and their interference, at the LHC with $\ecm = 14$~TeV for the top-only (a)-(c) and fourth-generation (d)-(f) scenarios.  These are calculated for the scalar mass $m_{h}=800$~GeV and for the phase $(\alpha / \pi)=0, 0.25, 0.50$ and have been integrated with PDFs in the range $790\gev<\rts<800\gev$.  The magnitude of the normalized scalar-top coupling $A$ is fixed at $|A|=1$.}
\label{fig:phi_800}
\end{figure}

We show a more realistic possibility in Figs.~\ref{fig:phi_800}(a)-\ref{fig:phi_800}(f).  This time we use a scalar mass of $m_{h}=800$~GeV and integrate over the region $790\gev<\rts<800\gev$ for the same reasons described above.  Not only are the signals less pronounced here in the invariant mass distribution, but the QCD contribution is roughly an order of magnitude smaller than in the previous case as well.  The distributions in Figs.~\ref{fig:phi_800}(a)-\ref{fig:phi_800}(f) are therefore more representative of possible signals that may currently be hidden.  The top-only case creates essentially no deviation from the QCD prediction, as expected from its lack of observable signal in the $M_{t\tbar}$ distribution.  However, in the fourth-generation case the deviations are small but seemingly distinguishable.  This is aided by the fact that at this center-of-mass energy, the $\Phi$ distribution predicted by QCD is nearly flat.  Also due to this fact, the phase of the total distribution very nearly matches that of the signal, with $\delta_{tot}\approx - 2 \chi \approx  -2 \alpha$, and the distributions for each of the three values of $\alpha$ shown are clearly distinguishable.

\begin{figure}
\centering
\subfigure[]{\includegraphics[clip,width=0.45\textwidth]{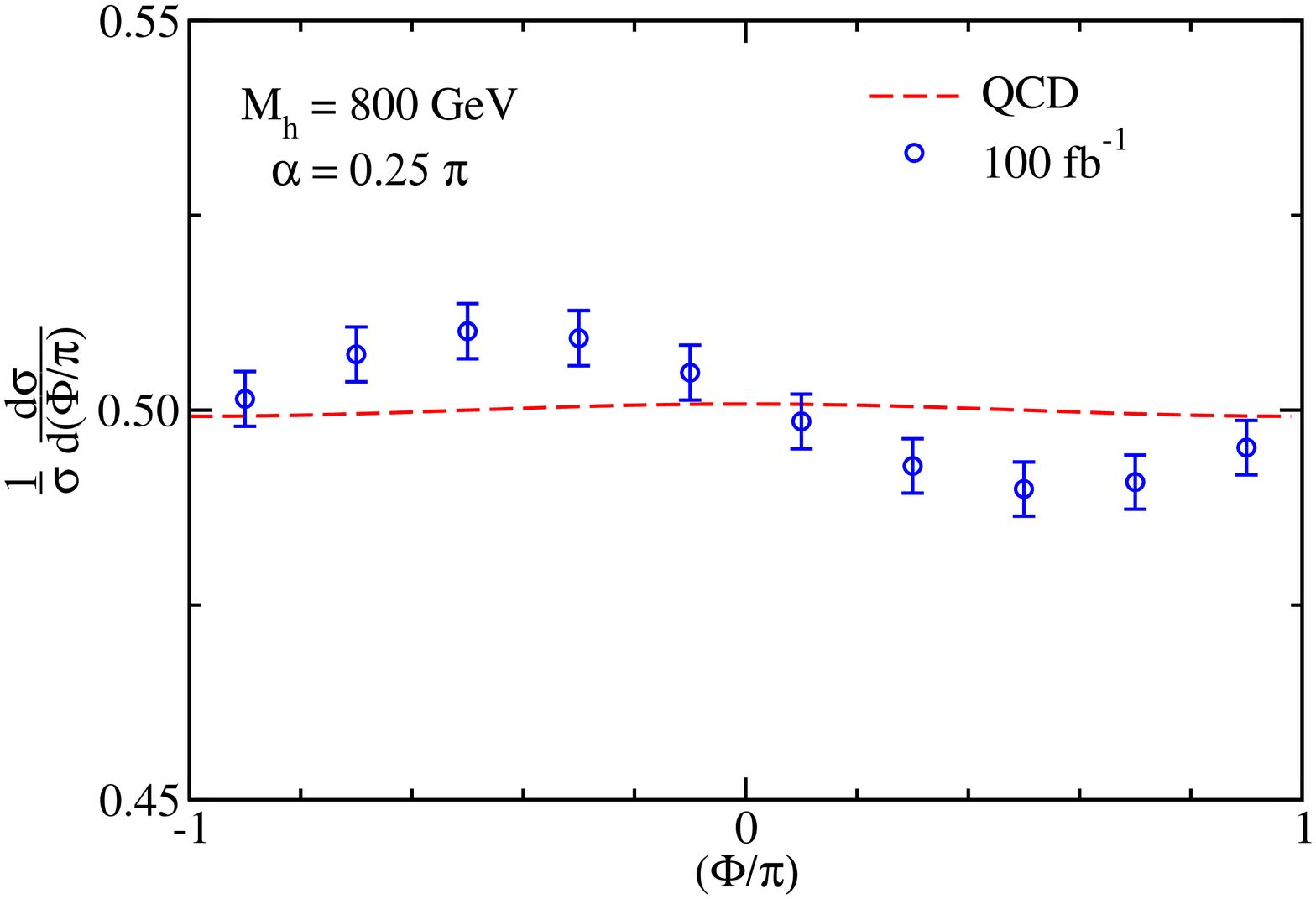}}
\subfigure[]{\includegraphics[clip,width=0.45\textwidth]{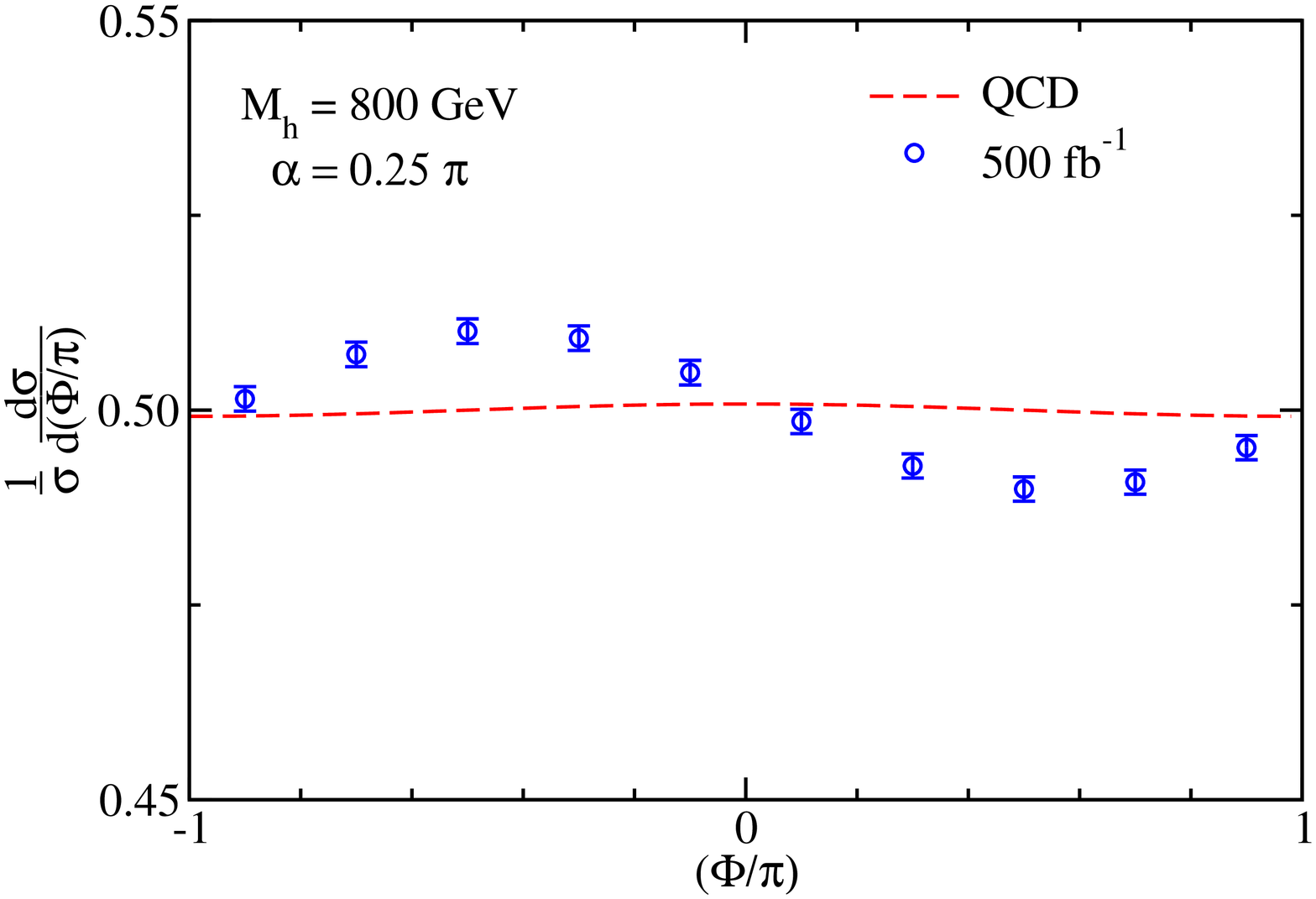}}
\caption{The data points are a "binned" version of the full expected $\Phi$ distribution in the region 790~GeV$< \rts <$800~GeV for a scalar mass $m_{h}=800$~GeV and phase $\alpha=0.25\pi$ at the LHC with $\ecm = 14$~TeV.  The error bars signify the expected magnitude of the statistical uncertainty in each bin for a luminosity of (a) 100 fb$^{-1}$ and (b) 500 fb$^{-1}$.  The dashed red curve gives the LO QCD background prediction for distribution from the process $gg \to t \tbar \to b \bbar \wplus \wminus$.}
\label{fig:stats}
\end{figure}

It is important to note that the size of the deviations described above are only meaningful when understood within the context of the relevant systematic and statistical uncertainties of the experiment.  While we do not perform a full Monte Carlo simulation, we attempt to get an estimate of the statistical uncertainties involved by plotting a "binned" version of the total normalized $\Phi$ distribution.  This is performed by integrating over regions of $\Delta(\Phi / \pi) = 0.2$, dividing by the bin size, and plotting the error bars that would be associated with the counting statistics on the number of events in each bin predicted for a given luminosity.  We do this in Figs.~\ref{fig:stats}(a) and \ref{fig:stats}(b) for integrated luminosities of $\mathscr{L}=100$~fb$^{-1}$ and $\mathscr{L}=500$~fb$^{-1}$, respectively, for the fourth-generation scenario with $m_{h}=800$~GeV and $\alpha=0.25\pi$.  The QCD prediction is plotted as a smooth curve.  What can be seen in the comparison between Figs.~\ref{fig:stats}(a) and \ref{fig:stats}(b) is that in this more realistic scenario, a very large amount of data (on the order of hundreds of fb$^{-1}$) will be necessary to begin to have sensitivity to the deviations at this level, but that they are in principle possible to observe.  For smaller scalar masses this requirement would be alleviated somewhat due to the more favorable statistics coming from more pronounced signals and larger overall rates.

We close with a few comments on additional theoretical concerns.  Firstly, it has been noted previously in Refs.~\cite{Frederix:2007gi,Mahlon:2010gw,Baumgart:2011wk} that the azimuthal angular correlations of the $W$-boson decay products exhibit a stronger modulating signal than those of the $W$ bosons themselves due to a 100\% spin correlation of the down-type decay objects with the polarization of the parent top quark.  While we have worked in the on-shell $W$ approximation primarily for ease of calculation, we expect little improvement from such an extension of the calculation, as both the signal and background will be affected and therefore the question of detecting a deviation remains.  A larger overall modulation could, however, improve the statistics necessary for detection and is therefore worth consideration.

Secondly, we note here that we have dealt only with the gluon initiated process and have ignored the processes $q\bar{q} \to t \tbar$.  This should be a fairly good approximation:  while $q\bar{q} \to t \tbar$ makes up a small but non-negligible fraction to the total $pp \to t \tbar$ cross section at $\ecm = 14$~TeV, it provides a continuum distribution much like the $gg\to t \tbar$ contribution and should therefore be insignificant when considering the small, isolated ranges of $M_{t\tbar}$ used here.

Finally, it is well known that higher-order contributions to the total $pp \to t \tbar$ cross section in QCD are quite large, resulting in $K$ corrections of $\sim 1.5$ \cite{Bernreuther:2008ju}.  These, of course, would have an effect on the $d\sigma / dM_{t\tbar}$ distribution of both the QCD process and the signal.  Any difference in the corrections between the two would change the relative weights used in the total angular distribution and therefore have an effect on the final results.  In addition, it is possible that there could be modifications to the normalized angular distributions themselves.  While all of these effects would be necessary for a more accurate understanding of any deviations that may be observed in these distributions, we believe the results here will serve as a good approximation: because most higher-order corrections are related to initial- and final-state radiation, and because these states are the same for the QCD contributions and the scalar signals described here, we would expect them to be of comparable magnitude and therefore approximately cancel out in the normalized distributions we consider.  In addition, the next-to-leading-order corrections have been calculated for the azimuthal angular distribution of the final-state leptons resulting from the $W$-boson decays in the QCD process and these have been shown to be small at the LHC \cite{Melnikov:2009dn}.

\section{Conclusions} \label{sec:conclusions}
We have presented the LO prediction for the distribution of the azimuthal angle between the decay planes of the top quarks in the process $gg \to t \tbar \to b \bbar \wplus \wminus$.  We have then sought to demonstrate the deviations that may arise in this distribution from possible New Physics contributions by specifically considering the case of a heavy scalar particle with a  coupling to the top quark such that it can contribute via an on-shell decay.  We investigated this possibility within the context of two separate scenarios: one in which this coupling to the top is the only relevant coupling of the scalar, and another, inspired by fourth-generation models, that includes couplings to two new, heavy quarks.  We have placed special emphasis on understanding the contribution to the total azimuthal distribution from the interference of the signal with QCD and on the possibility of $CP$-violating couplings of the scalar to the top quark.  We have shown that, in certain scenarios, large amounts of integrated luminosity may allow this distribution to be used to directly probe $\alpha$, the phase associated with these couplings.

\section{Acknowledgments}
This work was supported by U.S. Department of Energy Grants No.~DE-FG02-95ER40896 and No.~DE-FG02-84ER40173, as well as Spanish research Grants No.~FPA2010-20807 and No.~2009SGR502, the Consolider CPAN project, and the Wisconsin Alumni Research Foundation.

\newpage
\bibliographystyle{h-physrev}
\bibliography{ms.bbl}
\newpage

\end{document}